\documentclass[journal=jpcafh,manuscript=article]{achemso}

\usepackage{chemformula} 
\usepackage[T1]{fontenc} 

\usepackage{graphicx}
\usepackage{subcaption}
\usepackage{amsmath}
\usepackage{amssymb}
\usepackage{braket}
\usepackage{bm}

\author{Paul A. Johnson}
\email{paul.johnson@chm.ulaval.ca}
\affiliation[Universit\'e Laval]
{D\'{e}partement de chimie, Universit\'{e} Laval, Qu\'{e}bec, Qu\'{e}bec, Canada}

\title{Beyond a Richardson-Gaudin mean-field: Slater-Condon rules and perturbation theory}
\abbreviations{IR,NMR,UV}

\begin{document}




\begin{abstract}
Richardson-Gaudin states provide a basis of the Hilbert space for strongly correlated electrons. In this study, optimal expressions for the transition density matrix elements between Richardson-Gaudin states are obtained with a cost comparable with the corresponding reduced density matrix elements. Analogues of the Slater-Condon rules are identified based on the number of near-zero singular values of the RG state overlap matrix. Finally, a perturbative approach is shown to be close in quality to a configuration interaction of Richardson-Gaudin states while being feasible to compute.
\end{abstract}

\section{Introduction}
Weakly correlated electrons are well described by Kohn-Sham density functional theory (KS-DFT) and coupled-cluster (CC) theory.\cite{cizek:1966,cizek:1971,purvis_1982,bartlett_book,bartlett:2007,helgaker_book} The qualitative behaviour is approximated by a Slater determinant and corrected by an approximate functional (KS-DFT) or by developing the wavefunction (CC). Both approaches are succesful as the corrections are small. In either case the approach is well understood, systematic improvement is possible, and results are easily computable by the non-expert with software packages.

For strongly correlated electrons there is generally no obvious Slater determinant reference. Many Slater determinants are required so that a qualitative mean-field description is difficult. The complete active space self-consistent field (CASSCF)\cite{roos:1980,siegbahn:1980,siegbahn:1981,roos:1987} is successful if the important orbitals are few enough in number and easily identified. Approximate CASSCF solvers\cite{olsen:1988,malmqvist:1990,fleig:2001,ma:2011,manni:2013,thomas:2015,manni:2016,schriber:2016,levine:2020} have been developed to treat larger systems, for which the density matrix renormalization group (DMRG)\cite{chan:2002,ghosh:2008,yanai:2009,wouters:2014,sun:2017,ma:2017} has also been applied with great success. These methods target the wavefunction directly in a computationally efficient manner without making reference to a mean-field picture.

It is understood that pairs of electrons are a more appropriate starting point for strongly correlated electrons.\cite{fock:1950} The antisymmetrized product of interacting geminals (APIG) provides an excellent description of paired electrons for an intractable cost.\cite{silver:1969,silver:1970a,silver:1970b,silver:1970c,paldus:1972a,paldus:1972b,moisset:2022a} These results are obtainable at mean-field cost through the antisymmetric product of 1-reference orbital geminals (AP1roG),\cite{limacher:2013} or eqiuvalently, pair-coupled-cluster doubles (pCCD).\cite{stein:2014,limacher:2014a,limacher:2014b,henderson:2014a,henderson:2014b,boguslawski:2014a,boguslawski:2014b,boguslawski:2014c,tecmer:2014,boguslawski:2015,boguslawski:2016a,boguslawski:2016b,boguslawski:2017,boguslawski:2019,nowak:2019,boguslawski:2021,nowak:2021,marie:2021,kossoski:2021} AP1roG/pCCD shows excellent results for the ground state energy of systems with no unpaired electrons. Generalizations to systems with unpaired electrons, or broken pairs of electrons, is however not obvious.

The reduced Bardeen-Cooper-Schrieffer (BCS)\cite{bardeen:1957a,bardeen:1957b,schrieffer_book} Hamiltonian is an exactly solvable model whose eigenvectors are products of pairs of electrons. These states were first obtained by Richardson\cite{richardson:1963,richardson:1964,richardson:1965} and generalized by Gaudin\cite{gaudin:1976} and are thus referred to as Richardson-Gaudin (RG) states. As eigenvectors of a physical model, RG states form a basis for the Hilbert space. It has been shown for strongly correlated model systems that a single RG state is a good starting point, much like a Slater determinant for weakly correlated systems.\cite{johnson:2020,fecteau:2022,johnson:2023}

RG states are feasible as their reduced density matrix (RDM) elements are cheap to compute,\cite{gorohovsky:2011,fecteau:2020,faribault:2022} and their couplings with excited states decay rapidly with excitation level.\cite{johnson:2021} Previous constructions of the transition density matrix (TDM) elements to perform a configuration interaction with singles and doubles (CISD) of RG states scaled very poorly, which is a situation remedied herein. The purpose of this contribution is to nail down the details of using RG states for fully paired electrons, the so-called seniority-zero sector. This is meant as a stepping stone to RG states with unpaired electrons, which are much more complicated and will be treated in a future contribution.

This manuscript is organized as follows. First, a brief overview of RG states is given, discussing in particular their density matrix elements and low-lying excited states. Optimal expressions for the TDM elements are obtained using an approach developed by Chen and Scuseria for Wick's theorem applied to Hartree-Fock-Bogoliubov states.\cite{chen:2023} Second, contributions of the individual RG states to the RGCISD wavefunction are studied, leading to analogues of the Slater-Condon rules for RG states. Finally, perturbation theory is shown to give comparable results to RGCISD at a fraction of the cost.

\section{Richardson-Gaudin States}
This section provides a brief outline of Richardson-Gaudin states, keeping the details to a minimum. For a complete description see ref.\cite{johnson:2023b} There are \emph{always} $N$ sites or spatial orbitals labelled with indices $i,j,k,l$, and $M$ pairs labelled with indices $a,b$. 

RG states are built from pairs of electrons, which are elementary representations of the Lie algebra su(2). For each spatial orbital there are three operators
\begin{align} \label{eq:pair_su2}
	S^+_i = a^{\dagger}_{i\uparrow}a^{\dagger}_{i\downarrow}, 
	\quad S^-_i = a_{i\downarrow}a_{i\uparrow},
	\quad S^z_i = \frac{1}{2}\left( a^{\dagger}_{i\uparrow}a_{i\uparrow} + a^{\dagger}_{i\downarrow}a_{i\downarrow} -1 \right).
\end{align}
$S^+_i$ creates a pair in spatial orbital $i$, $S^-_i$ removes a pair in spatial orbital $i$, while $S^z_i$ gives $+\frac{1}{2}$ for a pair-occupied spatial orbital, $-\frac{1}{2}$ for an empty spatial orbital, and 0 for a singly-occupied orbital. These operators have the usual structure of the spin operators of su(2)
\begin{align} \label{eq:su2_structure}
	[S^+_i, S^-_j] &= 2\delta_{ij} S^z_i \\
	[S^z_i, S^{\pm}_j] &= \pm \delta_{ij} S^{\pm}_i.
\end{align}
It is convenient to use $\hat{n}_i = 2S^z_i +1$, which counts the number of electrons in spatial orbital $i$. RG states are the eigenvectors of the reduced BCS Hamiltonian
\begin{align} \label{eq:H_bcs}
	\hat{H}_{BCS} = \frac{1}{2} \sum^N_{k=1} \varepsilon_k \hat{n}_k -\frac{g}{2} \sum^N_{k,l=1} S^+_k S^-_l.
\end{align}
They are products of RG pairs
\begin{align} \label{eq:rg_pairs}
	S^+ (u) = \sum^N_{i=1} \frac{S^+_i}{u-\varepsilon_i}
\end{align}
which depend on a set of complex numbers usually called \emph{rapidities}. The RG pair \eqref{eq:rg_pairs} is a linear combination of all one-pair states weighted by the difference of the rapidity and the energy $\varepsilon_i$ of the spatial orbital $i$. Rapidities are not free variables. In particular for the RG state
\begin{align} \label{eq:rg_state}
	\ket{\{u\}} = S^+(u_1) S^+(u_2) \dots S^+(u_M) \ket{\theta}
\end{align}
to be an eigenvector of the reduced BCS Hamiltonian, the rapidities $\{u\}$ must be a solution of the nonlinear equations
\begin{align} \label{eq:rich}
	\frac{2}{g} + \sum^N_{i=1} \frac{1}{u_a - \varepsilon_i} + \sum^M_{b (\neq a)=1} \frac{2}{u_b - u_a} = 0, \qquad \forall a = 1,\dots M
\end{align}
first derived by Richardson, and are hence known as Richardson's equations. Each RG state is defined by a complete set of rapidities that together solve Richardson's equations. Thus, strictly speaking there are no rapidities shared between eigenvectors of the reduced BCS Hamiltonian. 

Numerically solving Richardson's equations is possible, but extra care must be taken near the critical points where a rapidity coincides with a single-particle energy.\cite{rombouts:2004,guan:2012,pogosov:2012,debaerdemacker:2012} It is far easier to work with so-called \emph{eigenvalue-based variables} (EBV)
\begin{align} \label{eq:ebv}
	U_i &= \sum^M_{a=1} \frac{g}{\varepsilon_i - u_a}
\end{align}
which are solutions of the non-linear equations
\begin{align} \label{eq:ebv_eq}
	U^2_i - 2U_i -g \sum^N_{k (\neq i)=1} \frac{U_k - U_i}{\varepsilon_k - \varepsilon_i} = 0, \quad \forall i=1,\dots,N.
\end{align}
Taking the sum of Richardson's equations, for each $a$, one finds that the EBV satisfy the sum rule
\begin{align} \label{eq:ebv_sum}
	\sum^N_{i=1} U_i = 2M.
\end{align}
These equations are easy to solve numerically and do not have troublesome critical points. For the complete details of solving the EBV equations \eqref{eq:ebv_eq} see refs.\cite{faribault:2011,elaraby:2012,fecteau:2022} It is convenient to use rapidities when looking at the geminal coefficients in each pair \eqref{eq:rg_pairs}, but numerically they are not productive to compute. RG pairs and states have clear definitions in terms of rapidities but do not in terms of EBV. Thus, we continue to label RG states with rapidities even if we do not compute them numerically.

When $g=0$, the reduced BCS Hamiltonian is non-interacting and the EBV equations \eqref{eq:ebv_eq} decouple completely
\begin{align} 
	U_i (U_i -2) = 0,
\end{align}
whose solutions are $M$ EBV equal to 2 and $N-M$ EBV equal to 0. These RG states are Slater determinants whose pair-occupied orbitals correspond to the particular EBV equal to 2. The remarkable fact is that as $g$ grows in either direction, the RG states are connected uniquely to one solution at $g=0$.\cite{yuz:2003,yuz:2005} RG states can thus be labelled by the bitstring of the Slater determinant they represent at $g=0$ even though at finite $g$ they \emph{are not} Slater determinants. The ground state of the reduced BCS Hamiltonian is always labelled by $M$ 1s followed by $N-M$ 0s, while the highest excited state is always labelled by $N-M$ 0s followed by $M$ 1s. Other RG states \emph{will} cross, but these are always strict crossings and not avoided crossings.\cite{yuz:2002,sklyanin:1989,cambiaggio:1997}

\subsection{Density matrix elements}
Expressions for the density matrices are known both in terms of rapidities\cite{faribault:2008,faribault:2010,gorohovsky:2011,fecteau:2020} and more recently in terms of the EBV.\cite{faribault:2022} These formulas require the rapidities to be solutions of Richardson's equations \eqref{eq:rich} or the EBV to be solutions of the EBV equations \eqref{eq:ebv_eq}: they are \emph{on-shell}. If the rapidities do not satisfy Richardson's equations, or the EBV do not satisfy the EBV equations \eqref{eq:ebv_eq}, they are \emph{off-shell}. The expressions in terms of the EBV are preferred as they avoid unnecessary computations and are more stable.

The physical systems we wish to solve are Coulomb Hamiltonians
\begin{align} \label{eq:C_ham}
	\hat{H}_C = \sum^N_{i,j=1} h_{ij} \sum_{\sigma} a^{\dagger}_{i \sigma} a_{j \sigma} + \frac{1}{2} \sum^N_{i,j,k,l=1} V_{ijkl} \sum_{\sigma \tau} a^{\dagger}_{i \sigma} a^{\dagger}_{j \tau} a_{l \tau} a_{k \sigma}
\end{align}
where the 1- and 2-electron integrals are expressed in a basis $\{\phi\}$
\begin{align}
	h_{ij} &= \int d\mathbf{r} \phi^*_i (\mathbf{r}) \left( - \frac{1}{2} \nabla^2 - \sum_I \frac{Z_I}{| \mathbf{r} - \mathbf{R}_I |} \right) \phi_j (\mathbf{r}) \\
	V_{ijkl} &= \int d\mathbf{r}_1 d\mathbf{r}_2 \frac{\phi^*_i(\mathbf{r}_1)  \phi^*_j(\mathbf{r}_2)  \phi_k(\mathbf{r}_1)  \phi_l(\mathbf{r}_2)  }{| \mathbf{r}_1 - \mathbf{r}_2|}.
\end{align}
$\sigma$ and $\tau$ represent spin labels, and the integrals are taken in a restricted formalism. Unrestricted and generalized integrals may be used with no formal complication. The RDM elements for RG states are simple. In particular, the energy expression for the Hamiltonian \eqref{eq:C_ham} evaluated with an RG state is 
\begin{align} \label{eq:sz_energy}
	E [\{\varepsilon\},g] = 2 \sum^N_{k=1} h_{kk} \gamma_k + \sum^N_{k=1} \sum^N_{l (\neq k) =1} (2V_{klkl} - V_{kllk})D_{kl} + \sum^N_{k,l=1} V_{kkll} P_{kl},
\end{align}
where the 1-RDM elements $\gamma_k$ are diagonal, while the 2-RDM reduces to 2 $N \times N$ matrices: the diagonal-correlation function $D_{kl}$ and the pair-correlation function $P_{kl}$
\begin{subequations} \label{eq:sz_dm}
	\begin{align}
		\gamma_k &= \frac{1}{2} \frac{\braket{ \{u\} | \hat{n}_k | \{u\} }}{\braket{ \{u\} | \{u\} }} \\
		D_{kl} &= \frac{1}{4}   \frac{\braket{ \{u\} | \hat{n}_k \hat{n}_l | \{u\} }} {\braket{ \{u\} | \{u\} }} \\
		P_{kl} &= \frac{\braket{\{u\} | S^+_k S^-_l | \{u\}}} {\braket{\{u\} | \{u\} }}.
	\end{align}
\end{subequations}
From the representation of the pair operators \eqref{eq:pair_su2}, it is seen that the diagonal elements $D_{kk}$ and $P_{kk}$ refer to the same element of the 2-RDM, which is itself equal to $\gamma_k$. Arbitrarily, this value is assigned to $P_{kk}=\gamma_k$ and $D_{kk}$ is set to zero. 

Scalar products between RG states have a simple expression in terms of a determinant. For $\{u\}$ on-shell, and $\{v\}$ arbitrary,
\begin{align} \label{eq:ebv_scalar}
	\braket{\{u\}|\{v\}} = \eta \det J
\end{align}
with the factor $\eta = (-1)^{N-M} \left( \frac{g}{2} \right)^{-2M}$ and the matrix
\begin{align}
	J_{kl} =
	\begin{cases}
		U_k + V_k - 2 + \sum^N_{i (\neq k)=1} \frac{g}{\varepsilon_i - \varepsilon_k}, \quad & k=l, \\
		\frac{g}{\varepsilon_k - \varepsilon_l}, \quad & k\neq l.
	\end{cases}
\end{align}
When normalized, the factor $\eta$ will appear both in the numerator and the denominator, so the factor $(\frac{g}{2})^{-2M}$ can be omitted to avoid overflows. It is important however to keep track of the sign. Henceforth, $\eta = (-1)^{N-M}$.

One- and two-body density matrix elements are computable from the first
\begin{align}
	A[J]^{i,k} = (-1)^{i+k} \eta \det J^{i,k}
\end{align}
 and second cofactors of $J$
\begin{align}
	A[J]^{ij,kl} = (-1)^{i+j+k+l+h(i-j)+h(k-l)} \eta \det J^{ij,kl}.
\end{align}
$J^{i,k}$ is $J$ without the $i$th row and the $k$th column, and $J^{ij,kl}$ is $J$ without the $i$th and $j$th rows, and the $k$th and $l$th columns. Second cofactors are antisymmetric with respect to the exchange of $i,j$ and $k,l$ which is accounted with the Heaviside function $h(x)$
\begin{align}
	h(x) = 
	\begin{cases}
		1 & x > 0 \\
		0 & x \leq 0.
	\end{cases}
\end{align}
The factor of $\eta$ has been absorbed into the cofactors to clean up the resulting expressions. The RDM and TDM expressions are equivalent, differing only in the way the cofactors are computed. The 1-body elements are a single summation over first cofactors
\begin{align}
	\gamma^{uv}_k := \frac{1}{2} \braket{\{u\}|\hat{n}_k|\{v\}} = \sum^N_{l=1} V_l A[J]^{l,k}.
\end{align}
With 
\begin{align}
	K_{kl} = V_k V_l + g\frac{V_k - V_l}{\varepsilon_k - \varepsilon_l},
\end{align}
2-body diagonal elements are summations over second cofactors

\begin{align}
	D^{uv}_{kl} :&= \frac{1}{4} \braket{\{u\}|\hat{n}_k \hat{n}_l |\{v\}} \\
	&= K_{kl} A[J]^{kl,kl} + \sum^N_{i (\neq k,l)=1} K_{il}A[J]^{il,kl} + \sum^N_{i(\neq k,l)=1} K_{ik}A[J]^{ki,kl} \nonumber \\
	&+ \sum^N_{i (\neq k,l)=1} \sum^N_{j (\neq k,l) = i +1} \frac{(\varepsilon_k-\varepsilon_i)(\varepsilon_l-\varepsilon_j) +(\varepsilon_k-\varepsilon_j)(\varepsilon_l-\varepsilon_i)}{(\varepsilon_k-\varepsilon_l)(\varepsilon_j-\varepsilon_i)} K_{ij} A[J]^{ij,kl},
\end{align}
while the pair-transfer elements are summations over first and second cofactors
\begin{align}
	P^{uv}_{kl} :&= \braket{\{u\}|S^+_k S^-_l | \{v\}} \\
	&= \left(V_l + \frac{(\varepsilon_k-\varepsilon_l)}{g}(V_l V_l - V_l J_{ll}) \right) A[J]^{l,k} + \sum^N_{i (\neq k,l)=1} \frac{\varepsilon_i - \varepsilon_k}{\varepsilon_i - \varepsilon_l} V_i A[J]^{i,k} \nonumber \\
	&- 2 \sum^N_{i (\neq k,l)=1} \frac{\varepsilon_k - \varepsilon_i}{\varepsilon_l - \varepsilon_i} K_{il} A[J]^{il,kl} 
	- 2 \sum^N_{i (\neq k,l)=1} \sum^N_{j (\neq k,l) = i +1} \frac{(\varepsilon_k-\varepsilon_i)(\varepsilon_k-\varepsilon_j)}{(\varepsilon_k-\varepsilon_l)(\varepsilon_j-\varepsilon_i)} K_{ij} A[J]^{ij,kl}.
\end{align}
Thus, to effectively compute the density matrix elements, the first and second cofactors of $J$ must be computed as efficiently as possible. 

When the two RG states are the same and on-shell $\{v\}=\{u\}$, the matrix $J$ becomes the Jacobian of the EBV equations, which is emphasized as $\bar{J}$
\begin{align}
	\bar{J}_{kl} =
	\begin{cases}
		2 U_k - 2 + \sum^N_{i (\neq k)=1} \frac{g}{\varepsilon_i - \varepsilon_k}, \quad & k=l, \\
		\frac{g}{\varepsilon_k - \varepsilon_l}, \quad & k\neq l.
	\end{cases}
\end{align}
As $\bar{J}$'s determinant is the square of the norm of a physical state, it is well-conditioned and can be inverted. (There are exceptions when the $\{\varepsilon\}$ become exactly degenerate. This is an issue that will be treated in a separate contribution.) The adjugate formula for the inverse ensures that the elements of $\bar{J}^{-1}$ are the normalized first cofactors of $\bar{J}$
\begin{align} \label{eq:1st_cof}
	\bar{J}^{-1}_{ij} = \frac{A[J]^{j,i}}{\eta \det \bar{J}},
\end{align}
which are the actual quantities desired to calculate RDM elements. Normalized second cofactors are obtained from a theorem of Jacobi\cite{vein_book}
\begin{align} \label{eq:2nd_cof}
	\frac{A[\bar{J}]^{ij,kl}}{\eta \det \bar{J}} &=
	  \frac{A[\bar{J}]^{i,k}}{\eta \det \bar{J}} \frac{A[\bar{J}]^{j,l}}{\eta \det \bar{J}}
	- \frac{A[\bar{J}]^{i,l}}{\eta \det \bar{J}} \frac{A[\bar{J}]^{j,k}}{\eta \det \bar{J}} 
	= 	\begin{vmatrix}
			\bar{J}^{-1}_{ki} & \bar{J}^{-1}_{li} \\
			\bar{J}^{-1}_{kj} & \bar{J}^{-1}_{lj}
		\end{vmatrix}
\end{align}
which holds so long as the matrix $\bar{J}$ is non-singular. If $p$th-order cofactors of $\bar{J}$ are required, Jacobi's theorem reduces them to a $p\times p$ determinant of the normalized first cofactors. Thus, to compute RDM elements, a single matrix inversion is required. 

When the two RG states are on-shell but distinct $\{v\} \neq \{u\}$, the determinant of $J$ must be zero as it represents the overlap of different non-degenerate eigenvectors of the reduced BCS Hamiltonian. The linear combination of the columns
\begin{align}
	\sum_j (U_j - V_j) J_{ij} = 0
\end{align}
is zero, and thus the matrix $J$ has rank $N-1$.\cite{claeys:2017b} As a result, the approach of inverting $J$ and using Jacobi's theorem must be modified. The way to proceed appeared recently in a related but different context.\cite{chen:2023} Managing the divergences is possible through the singular value decomposition (SVD) of $J$
\begin{align}
	J = \mathcal{U} \varSigma \mathcal{V}^{\dagger}.
\end{align}
The matrices $\mathcal{U}$ and $\mathcal{V}$ should not be confused with the EBV \eqref{eq:ebv}, they are the usual unitary matrices defining the SVD. Formally, $J$ has rank $N-1$, so has one singular value, $\sigma_N$ that is very small. It won't be exactly zero as the EBV are computed as the solutions of the EBV equations \eqref{eq:ebv_eq} to double precision. As $\sigma_N$ is present regardless of the RG states involved, it will be referred to as the fundamental singular value. Additional small singular values will appear based on which states are involved, but are generally orders of magnitude larger than $\sigma_N$.

The strategy to evaluate the cofactors is now straightforward. The inverse, computed by the SVD, has poles corresponding to the small singular values: $J^{-1}$ may be separated into a singular part $J^{\mathcal{S}}$ and a regular part $J^{\mathcal{R}}$
\begin{align}
	J^{-1} &= \det \mathcal{U} \det \mathcal{V} \prod_{\beta} \sigma_{\beta} \sum_{\alpha} \frac{1}{\sigma_{\alpha}} \mathcal{V}_{\alpha} \mathcal{U}^{\dagger}_{\alpha}
		   = J^{\mathcal{S}} + J^{\mathcal{R}}.
\end{align}
The individual singular values $\sigma_{\alpha}$ near zero are grouped into the set $\mathcal{S}$. Define $\zeta = \det \mathcal{U} \det \mathcal{V}$ and
\begin{align}
	\lambda^{\mathcal{R}} &= \prod_{\alpha \notin \mathcal{S}} \sigma_{\alpha} \\
	\lambda^{\mathcal{S}} &= \prod_{\alpha \in \mathcal{S}} \sigma_{\alpha}
\end{align}
so that $\det J = \zeta \lambda^{\mathcal{R}} \lambda^{\mathcal{S}}$. Further, the products of small singular values, without particular ones are 
\begin{align}
	\lambda^{\mathcal{S}}_{k} &= \prod_{\alpha \in \mathcal{S} \neq k} \sigma_{\alpha} \\
	\lambda^{\mathcal{S}}_{kl} &= \prod_{\alpha \in \mathcal{S} \neq k,l} \sigma_{\alpha}.
\end{align}

For small singular values $\alpha \in \mathcal{S}$ define
\begin{align}
	J^{\alpha} = \mathcal{V}_{\alpha} \mathcal{U}^{\dagger}_{\alpha}
\end{align}
so that the singular part of $J^{-1}$ is the sum
\begin{align}
	J^{\mathcal{S}} = \sum_{\alpha \in \mathcal{S}} \frac{1}{\sigma_{\alpha}} J^{\alpha}
\end{align}
with $\mathcal{U}_{\alpha}$ and $\mathcal{V}_{\alpha}$ the $\alpha$th columns of $\mathcal{U}$ and $\mathcal{V}$. The regular part of $J^{-1}$ is
\begin{align}
	J^{\mathcal{R}} = \sum_{\alpha \notin \mathcal{S}} \frac{1}{\sigma_{\alpha}} \mathcal{V}_{\alpha} \mathcal{U}^{\dagger}_{\alpha}.
\end{align}
The first cofactors can now be computed from \eqref{eq:1st_cof},
\begin{align} \label{eq:1st_cof_tdm}
	A[J]^{i,k} = \eta \det J J^{-1}_{ki} = (-1)^{N-M} \zeta \lambda^{\mathcal{R}} \sum_{\alpha \in \mathcal{S}} \lambda^{\mathcal{S}}_{\alpha} J^{\alpha}_{ik}.
\end{align}
The regular part $J^{\mathcal{R}}$ does not contribute as it will be scaled by all the singular values, and is thus zero. Likewise, the second cofactors are
\begin{align} \label{eq:2nd_cof_tdm}
	A[J]^{ij,kl} &= (-1)^{N-M}\zeta \lambda^{\mathcal{R}} (  \lambda^{\mathcal{S}} 
	(J^{\mathcal{R}}_{ik}J^{\mathcal{R}}_{jl} - J^{\mathcal{R}}_{il}J^{\mathcal{R}}_{jk}) 
	+ \sum_{\alpha \in \mathcal{S}} \lambda^{\mathcal{S}}_{\alpha} (J^{\alpha}_{ik} J^{\mathcal{R}}_{jl} + J^{\alpha}_{jl} J^{\mathcal{R}}_{ik}
	- J^{\alpha}_{il} J^{\mathcal{R}}_{jk} - J^{\alpha}_{jk} J^{\mathcal{R}}_{il} ) \nonumber \\
	+  &\sum_{\alpha < \beta \in \mathcal{S}} \lambda^S_{\alpha\beta} 
	( J^{\alpha}_{ik} J^{\beta}_{jl} + J^{\alpha}_{jl} J^{\beta}_{ik}
	- J^{\alpha}_{il} J^{\beta}_{jk} - J^{\alpha}_{jk} J^{\beta}_{il} ) )
\end{align}
Notice that in the double sum over the small singular values, there are no diagonal terms as these cancel exactly. The poles in the inverse have thus been removed and the construction may proceed. The cofactors are then normalized by dividing by the squareroot of the products of the norms of the two states. Thus, to compute the second cofactors we take the SVD of the matrix $J$, and construct the regular $J^R$ and singular $J^{\alpha}$ parts of its inverse. Usually there is only the single small singular value $\sigma_N$ to deal with, and this approach is very efficient: the first and second cofactors become
\begin{align}
	A[J]^{i,k} &= (-1)^{N-M} \zeta \lambda^{\mathcal{R}} J^N_{ik} \\
	A[J]^{ij,kl} &= (-1)^{N-M} \zeta \lambda^{\mathcal{R}}
	(J^N_{ik}J^{\mathcal{R}}_{jl} + J^N_{jl}J^{\mathcal{R}}_{ik} 
	-J^N_{il}J^{\mathcal{R}}_{jk} - J^N_{jk}J^{\mathcal{R}}_{il}).
\end{align}
Generally, there will be more than one small singular value, and a decision must be made regarding how small is to be considered small enough to be included in $\mathcal{S}$. In this contribution, if the ratio of the largest to the smallest singular values is at least $10^{8}$, the last singular value is removed and added to $\mathcal{S}$. This process is repeated until the ratio is smaller than $10^{8}$. Varying this threshold did not change the results at all. In such cases precision in the result is already lost.

Finally, notice that the number of small singular values in $J$ dictates the order of non-zero cofactors to be considered. In particular, if $J$ has two small singular values then $J$ has rank $N-2$ and its first cofactors vanish, while if $J$ has three small singular values then $J$ has rank $N-3$ and its second cofactors vanish etc. This has a direct analogue in the usual Wick's theorem for fermions and the Slater-Condon rules for Slater determinants. There, the number of zero singular values of the overlap matrix corresponds to the relative level of excitation of the two Slater determinants. Overlap matrices with more than two zero singular values give no 1- or 2-body transition elements. The distinction here is that the number of small singular values of $J$ can change as a function of the parameters $\{\varepsilon\}$ and $g$. It will be shown that there are patterns giving analogues of Slater-Condon rules.

Thus, for each pair of states, the cost of evaluating the TDM elements is now the same as the RDM elements for each individual state: there are $\mathcal{O}(N^2)$ elements to compute, and each requires a double sum over the second cofactors (which are computed on-the-fly from primitive summands) giving a scaling of $\mathcal{O}(N^4)$. Efficient grouping of the summations as matrix-vector products might reduce this cost by one order of magnitude.

\subsection{RG reference and excitations}
The results of the previous section hold for any RG state. The focus will now be drawn to those of chemical interest. Consider H$_2$ in a minimal basis, treated variationally with an RG state. This treatment is exact.\cite{johnson:2020} The reduced BCS Hamiltonian for a pair of electrons in two orbitals has three parameters: $\varepsilon_1$, $\varepsilon_2$ and $g$, though as two can be chosen to define the energy scale and energy reference point, there is a single degree of freedom. Thus, a variational treatment of minimal basis H$_2$ with an RG state is a one-variable problem, in particular the ratio $\frac{\varepsilon_2 - \varepsilon_1}{|g|}$. 
\begin{figure} [ht!]
	\begin{subfigure}{\textwidth}
		\centering
		\includegraphics[width=0.49\textwidth]{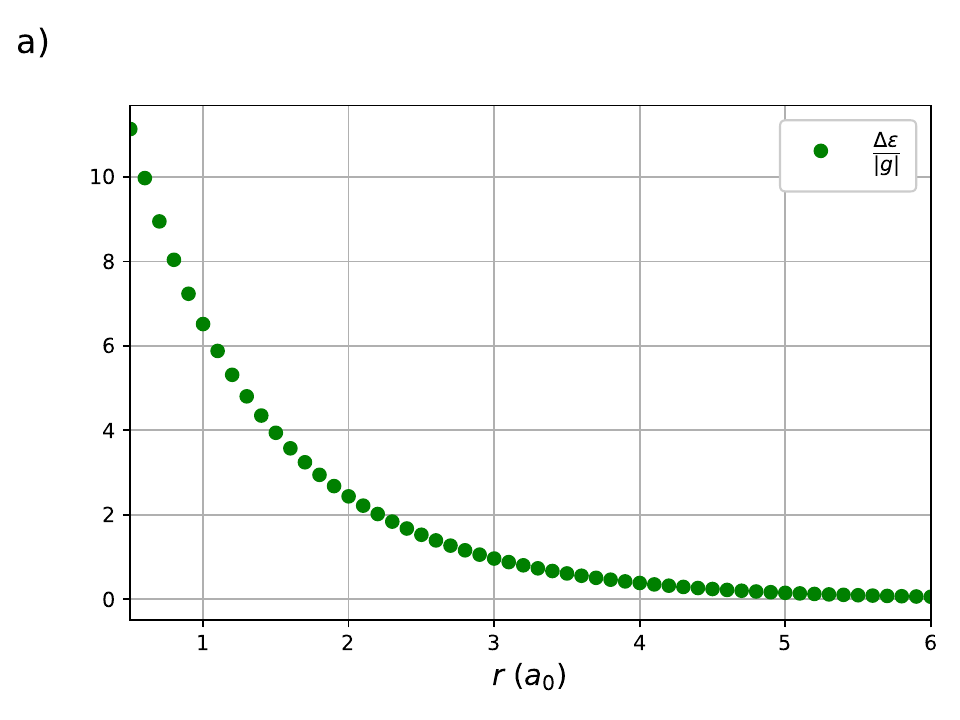} \hfill
		\includegraphics[width=0.49\textwidth]{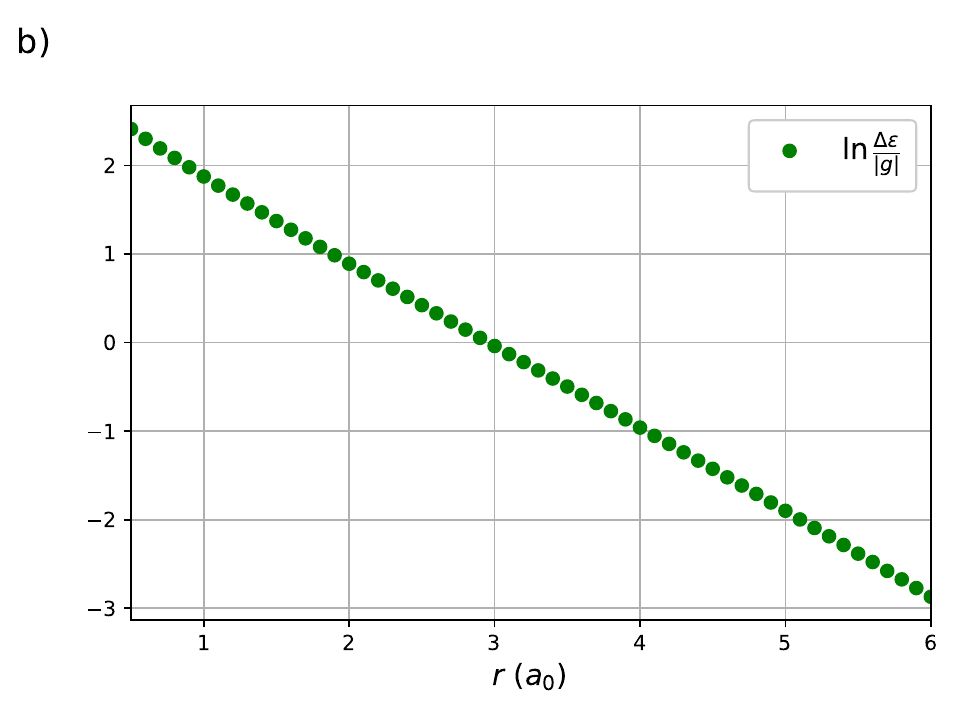}
	\end{subfigure}
	\caption{Plots of the single variational parameter, $\frac{\Delta \varepsilon}{|g|}$ as a function of $r$ for H$_2$ in the STO-6G basis set. (a) Linear plot. (b) Plot of natural logarithm. Least squares regression: $R^2=0.9998$, $r_C=2.97$ bohr.}
	\label{fig:h2_params}
\end{figure}
One sees in Figure \ref{fig:h2_params} (a) that this ratio appears to decay exponentially with the H -- H bond-length, which is confirmed in Figure \ref{fig:h2_params} (b) as $\ln \frac{\Delta \varepsilon}{|g|}$ is linear with $r$. The x-intercept, $r_C$, occurs where $\ln \frac{\Delta \varepsilon}{|g|} = 0$ or $\Delta \varepsilon = |g|$. For $r < r_C$, the orbital energy gap $\Delta \varepsilon$ is larger than the effective Coulomb repulsion $g$ and the effects of weak correlation dominate while the opposite is true when $r > r_C$. At $r_C$ the strong and weak correlation effects are in balance, and on either side of $r_C$ one of these effects decays rapidly.

In a previous report, it was found that the variationally optimal RG state for a 1D chain of $2M$ Hydrogen atoms is just $M$ copies of the 1-pair solution of H$_2$: the parameters $\varepsilon$ group into sets of two, and the corresponding RG pairs each become localized in two spatial orbitals.\cite{fecteau:2022} The corresponding reduced BCS Hamiltonian reduces to a valence-bond (VB) form
\begin{align} \label{eq:h_vb}
	\hat{H}_{VB} = \frac{1}{2} \sum^{M}_{a=1} (a-1) \xi \hat{n}_{2a} + ((a-1) \xi + \Delta \varepsilon_a) \hat{n}_{2a+1} - \frac{g}{2} \sum^N_{k,l=1} S^+_k S^-_l,
\end{align}
in terms of a large energy gap $\xi$, and small energy gaps $\Delta \varepsilon_a$ for each pair. The effective degrees of freedom have been substantially reduced, and the behaviour of the model is simple. The pairing strength $g$ is similar in size to $\Delta \varepsilon_a$ but is much smaller than $\xi$ resulting in a collection of nearly decoupled two-level subsystems, which will be named \emph{valence-bond subsystems} (VBS). Each VBS is denoted by indices $a=1,\dots,M$ and it is convenient to label their single particle energies $\varepsilon_{a_1}$ and $\varepsilon_{a_2}$ such that
\begin{align}
	\Delta \varepsilon_a = \varepsilon_{a_2} - \varepsilon_{a_1}.
\end{align}

The desired RG state places one pair in each VBS, which is the state labelled by the bitstring $(10)^M$ for $M$ pairs. When $g=0$, this state is a Slater determinant of the lower levels in each VBS, but any finite $g$ will cause partial occupation of both levels. Here, $\hat{H}_{VB}$ is for a half-filled valence system with no core or virtual levels. In general, there will be individual single-particle energies $\varepsilon$ that are isolated from one another compared with $g$ and are much lower (core) or higher (virtual) in energy than the $\varepsilon$ describing the valence-bonds. For $M_c$ pairs in the core and $M_v$ pairs in the valence, such that $M_c + M + M_v = N$, the desired RG state is $1^{M_c}(10)^M 0^{M_v}$. 
	
Each RG pair localizes in one VBS
\begin{align}
	S^+(u_a) = \sum_i \frac{S^+_i}{u_a - \varepsilon_i} \approx c_1 (u_a) S^+_{a_1} + c_2 (u_a) S^+_{a_2}
\end{align}
with small contributions from the other sites. This is the form of the generalized valence-bond / perfect pairing (GVB-PP)\cite{goddard:1967,hay:1972,hunt:1972,goddard:1973,goddard:1978,beran:2005,small:2009} wavefunction. It is not the ground state of $\hat{H}_{VB}$, but that doesn't matter: the parameters $\Delta \varepsilon_a$ and $g$ are auxiliary variables to describe the pairs, and do not represent physical energies. $\frac{\Delta \varepsilon_a}{g}$ again decays exponentially with the interatomic distance, though distinct pairs generally have different values of $r_C$. For linear H$_4$, both pairs have $r_C=3.1$ bohr, while in linear H$_8$ two of the pairs have $r_C=3.1$ bohr and the other two have $r_C=3.3$ bohr.\cite{johnson:2023b} 

The $(10)^M$ RG state, which henceforth will be denoted $\ket{M}$, represents $M$ pairs of electrons similar to GVB-PP. Like GVB-PP, correlation within each pair is well accounted while correlation between pairs is not. Unlike GVB-PP, $\ket{M}$ is an eigenvector of a model Hamiltonian, whose weak excitations can be constructed systematically to account for the missing interpair correlation. Single pair excitations have bitstrings that differ from the reference by one 1 and one 0, which can happen in two ways. Excitations within a VBS, \emph{swaps}, have bitstrings composed of $M-1$ (10)s and one (01). The $M$ swaps will be labelled $\ket{M^a_a}$, with $a$ denoting the VBS in which the pair has been excited. Excitations from one VBS to another, \emph{transfers}, have bitstrings of $M-2$ (10)s, one (00) and one (11). These will be labelled $\ket{M^a_b}$ to denote that a pair has been transferred from VBS $b$ to VBS $a$. 

Double excitations occur in 3 types based on how many indices are shared. There are double swaps $\ket{M^{ab}_{ab}}$, swap plus transfers $\ket{M^{ab}_{ad}}$, and double transfers $\ket{M^{ab}_{cd}}$. One might ask if this nomenclature is correct as a double swap could also be thought of as a double transfer $a\rightarrow b$ and $b\rightarrow a$. In reality, both of these processes are counted and it will be seen that these states are different from the other double transfers. Higher excitations can be labelled and classified in the same manner, but as their contributions are incredibly weak they will not be discussed further.

With the relevant RG states identified, their individual contributions will now be studied.

\section{Slater-Condon rules} \label{sec:selection_rules}
In ref.\cite{johnson:2021} it was seen that the RG ground state $1^M 0^{N-M}$ coupled appreciably only with its single and double pair excitations. This will now be established and justified for the RG state $\ket{M}$. Previous studies of small strongly correlated systems of Hydrogen atoms were very well described with a configuration interaction (CI) of RG states.\cite{johnson:2023,johnson:2023b} These systems do not exist in reality: a linear chain of equidistant Hydrogen atoms would undergo a Peirls distortion and become a set of independent H$_2$ molecules. However, these systems are studied as they maximize the effects of strong correlation while being small enough to treat exactly. 

RG states have no unpaired electrons: they have zero \emph{seniority}. They are thus approximations to a CI of all Slater determinants with no unpaired electrons, the so-called \emph{doubly-occupied configuration interaction} (DOCI). DOCI is not exact though it has been shown that classifying Slater determinants by seniority leads provides a systematic hierarchy for strongly correlated systems, even multiple bond-breaking processes like N$_2$.\cite{bytautas:2011} DOCI depends on the choice of orbitals: a Slater determinant that has no unpaired electrons in a given set of orbitals will have unpaired electrons in a different set of orbitals. It is therefore necessary to use orbital-optimized (OO-)DOCI. The OO-DOCI results for equidistant linear H$_4$ and H$_8$ were computed in the basis STO-6G in ref.\cite{johnson:2020} For these systems, a CI of the RG state $\ket{M}$ along with its singles and doubles, RGCISD, is numerically indistinguishable from OO-DOCI.\cite{johnson:2023b} The CI coefficients of each RG state contributing to RGCISD will now be studied to demonstrate that the expansion is short, and that the couplings between the RG states follow simple rules. 

\begin{figure} [ht!]
	\begin{subfigure}{\textwidth}
		\centering
		\includegraphics[width=0.49\textwidth]{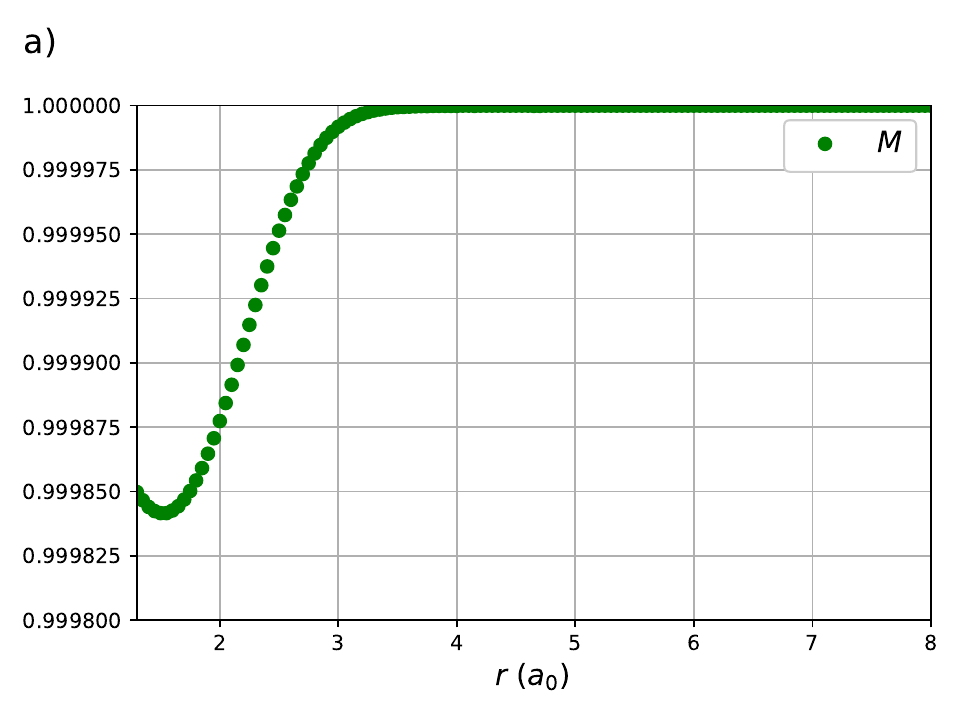} \hfill
		\includegraphics[width=0.49\textwidth]{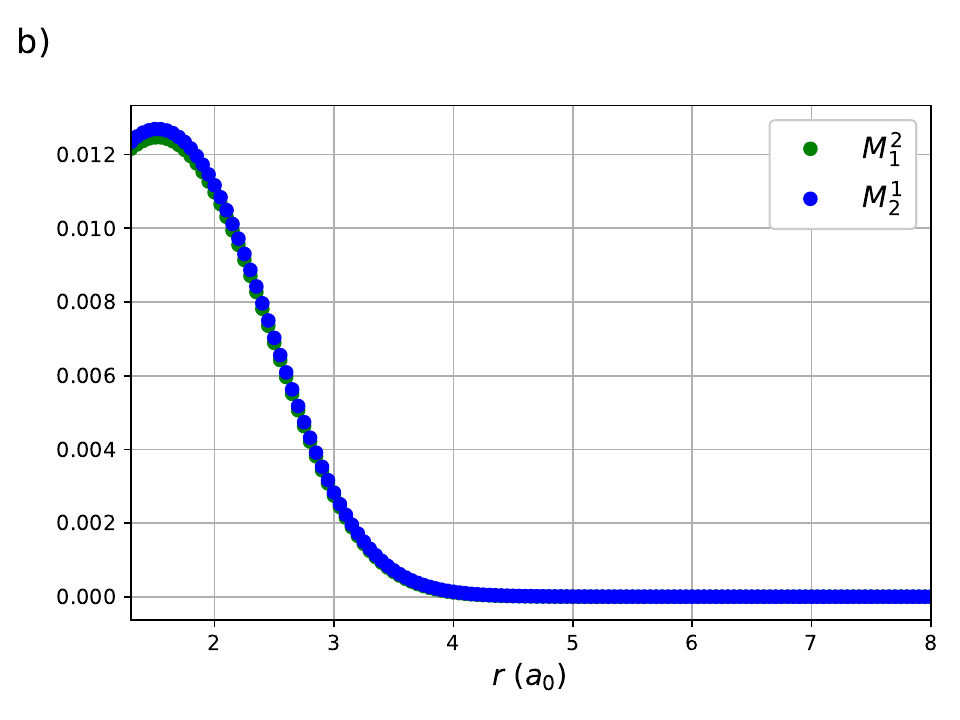} \\
		\includegraphics[width=0.49\textwidth]{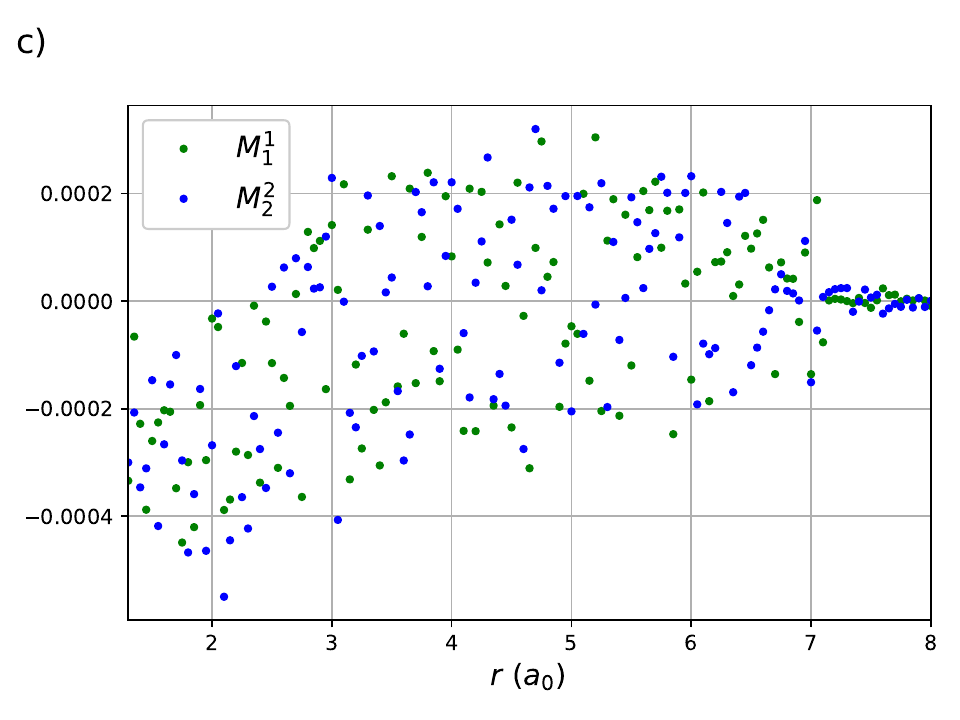} \hfill
		\includegraphics[width=0.49\textwidth]{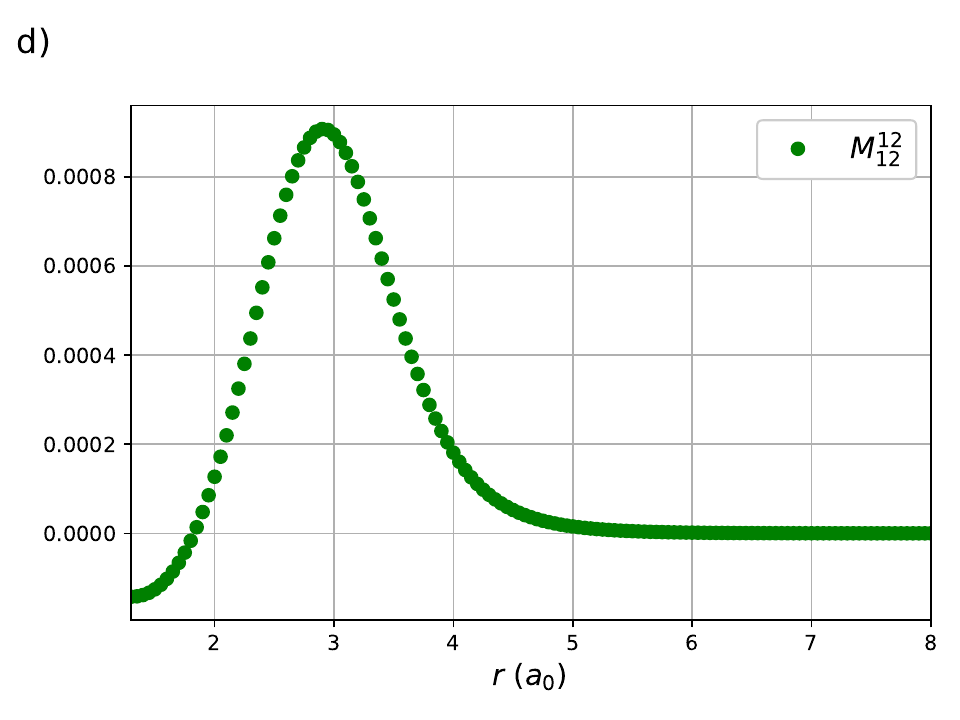}
	\end{subfigure}
	\caption{CI coefficients for the RGCISD ground state of linear H$_4$: (a) Reference RG state $\ket{M}$ (absolute value). (b) Single pair transfer states $\ket{M^{1}_{2}}$ and $\ket{M^{2}_{1}}$. (c) Single pair swap states $\ket{M^{1}_{1}}$ and $\ket{M^{2}_{2}}$. (d) Double pair swap state $\ket{M^{12}_{12}}$. Results computed with the OO-DOCI orbitals in the STO-6G basis set.}
	\label{fig:h4_states}
\end{figure}
CI coefficients of the RG states contributing to the ground state of equidistant linear H$_4$ are shown as a function of the H -- H distance in Figure \ref{fig:h4_states}. For this system, RGCISD and DOCI have the same cardinality and thus the two treatments are equivalent. Raw energy curves are not informative, as RGCISD is indistinguishable from OO-DOCI, but are shown in ref.\cite{johnson:2023b} As expected, the coefficient of the reference $\ket{M}$ completely dominates the expansion, only differing from 1 near the ``equilibrium'' geometry at $r=1.65$ bohr. The transfers $\ket{M^{2}_{1}}$ and $\ket{M^{1}_{2}}$ provide the interpair weak correlation and drop off very quickly once $r$ is greater than $r_C = 3.1$ bohr. The contributions from the swaps $\ket{M^{1}_{1}}$ and $\ket{M^{2}_{2}}$ are weak and centred at zero. Finally, the double swap $\ket{M^{12}_{12}}$ gives a weak contribution near equilibrium, but grows and peaks at $r=2.95$ bohr, which is just before $r_C$, before decaying back to zero.

\begin{figure} [ht!]
	\begin{subfigure}{\textwidth}
		\centering
		\includegraphics[width=0.49\textwidth]{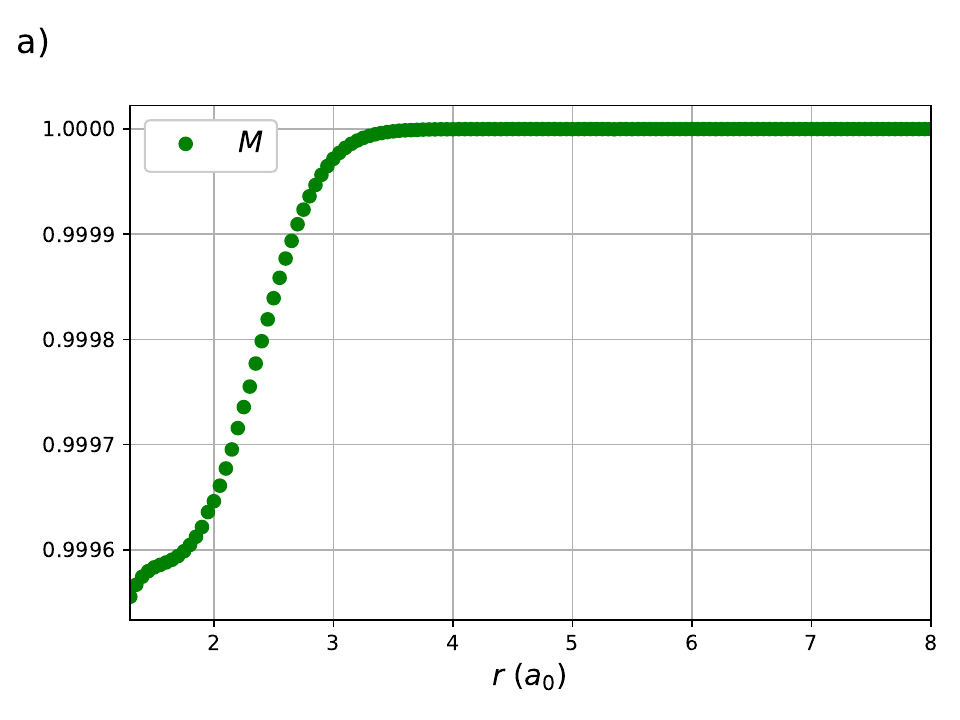} \hfill
		\includegraphics[width=0.49\textwidth]{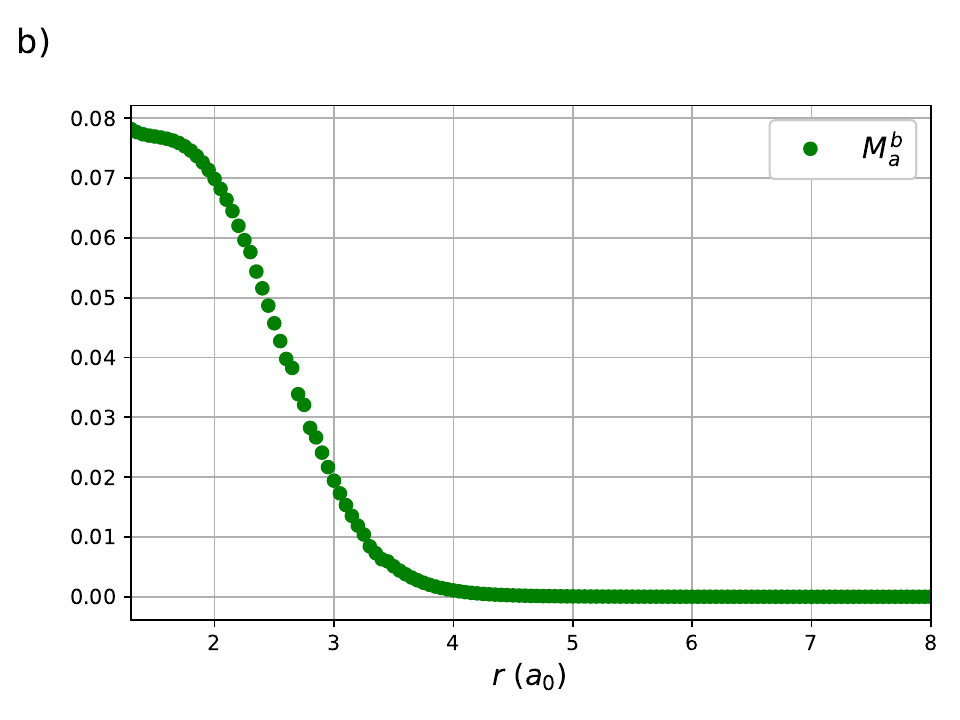} \\
		\includegraphics[width=0.49\textwidth]{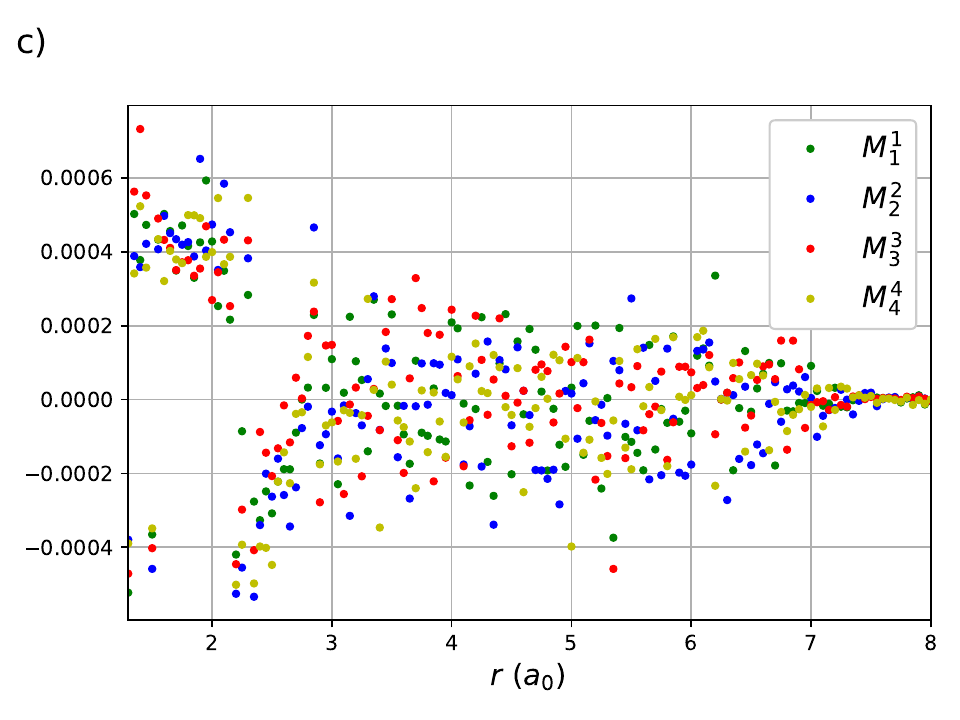} \hfill
		\includegraphics[width=0.49\textwidth]{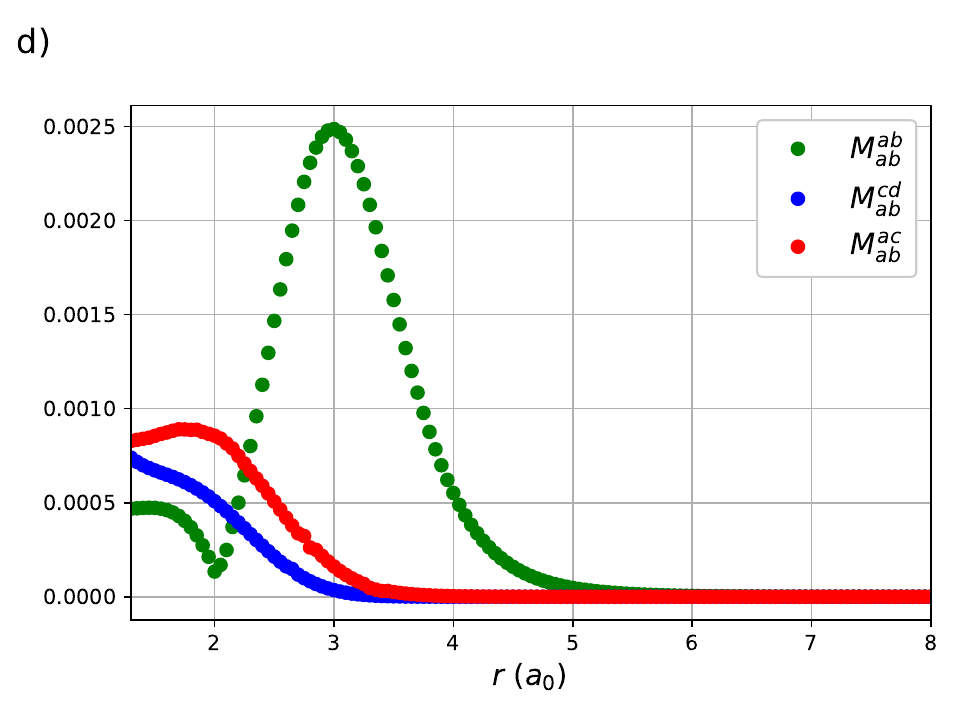}
	\end{subfigure}
	\caption{CI coefficients for the RGCISD ground state of linear H$_8$: (a) Reference RG state $\ket{M}$ (absolute value). (b) Sum of absolute values for single pair transfer states. (c) Single pair swap states. (d) Sum of absolute values for double pair excitations. Results computed with the OO-DOCI orbitals in the STO-6G basis set.}
	\label{fig:h8_states}
\end{figure}
The same behaviour is seen in the CI coefficients of the RG states for linear H$_8$. Here RGCISD is not DOCI, but it has been seen previously that the difference is on the order of $10^{-10}\;E_h$.\cite{johnson:2023b} Sums of absolute values of CI coefficients are plotted rather than individual coefficients for two reasons. First, to keep the plots clean and legible. Second, it is near-impossible to get clean continuous curves as even a very small discontinuity in any of the parameters, in particular the orbital coefficients, will cause a large difference in the individual CI coeffcients but no difference in the energy. Sums of absolute values of CI coefficients are invariant to such discontinuities while retaining the physical meaning. The expansion is completely dominated by the reference $\ket{M}$ with important contributions from the single transfers $\ket{M^{a}_{b}}$ near the equilibrium geometry at $r=1.75$ bohr. These contributions drop off very quickly, reaching roughly one tenth of their maximal value by $r=3.3$ bohr. The single swaps $\ket{M^{a}_{a}}$ are small, and on either side of zero. Individual coefficients for the swaps are plotted to convey the message. Contributions from the double swaps $\ket{M^{ab}_{ab}}$ are maximal at $r=3.0$ bohr, again just before the critical points $r_C$. 

The coupling of the reference $\ket{M}$ to each state goes to zero as $r$ becomes large, but in different ways. One might expect that all the TDM elements go to zero, but this is \emph{not} the case. Couplings between the reference and swaps have non-zero $\gamma_k$ and $P_{kl}$ elements within a VBS, but with opposite sign. In the large $r$ limit, the intra-VBS 1-body integrals $h_{a_1 a_1}= h_{a_2 a_2}$ are degenerate, as are the 2-body direct integrals
\begin{align}
	V_{a_1 a_1 a_1 a_1} = V_{a_1 a_1 a_2 a_2} = V_{a_2 a_2 a_1 a_1} = V_{a_2 a_2 a_2 a_2},
\end{align}
while the exchange integrals are always the same by symmetry
\begin{align}
	V_{a_1 a_2 a_1 a_2} = V_{a_2 a_1 a_2 a_1},
\end{align}
so the sum of $\gamma_k$ and $P_{kl}$ contributions will vanish. There are non-zero $D_{kl}$ elements between VBS, which at dissociation become for the particular VBS $a$ and all other VBS $b$
\begin{align}
	D_{b_1 a_1} = D_{b_2 a_1} = - D_{b_1 a_2} = - D_{b_2 a_2} = \frac{1}{4}.
\end{align}
The exchange integrals $V_{ijji}$ between VBS are zero as the orbitals are localized, and the direct elements are degenerate
\begin{align}
	V_{a_1 a_1 b_1 b_1} = V_{a_1 a_1 b_2 b_2} = V_{a_2 a_2 b_1 b_1} = V_{a_2 a_2 b_2 b_2}
\end{align}
so the couplings $\braket{M | \hat{H}_C | M^a_a}=0$ when $r$ becomes large.

Couplings between the reference and single transfers have no contribution from one-body elements past $r_C$. The TDM elements $\gamma_k$ themselves vanish as $J$ develops a second small singular value. While small, it is orders of magnitude larger than the fundamental singular value $\sigma_N$. At large $r$ there are small, but non-zero, $D_{kl}$ elements between VBS, i.e. for the coupling of the reference with the single transfer state $\ket{M^{b}_{a}}$ describing the pair transfer $a\rightarrow b$
\begin{align}
	D_{a_1 b_1} = - D_{a_1 b_2} = D_{b_1 a_1} = -D_{b_1 a_2} = -D_{b_2 a_1} = D_{b_2 a_2} = - D_{a_2 b_1} = D_{a_2 b_2}.
\end{align}
The direct integrals between VBS are degenerate
\begin{align}
	V_{a_1 a_1 b_1 b_1} = V_{a_1 a_1 b_2 b_2} = V_{a_2 a_2 b_1 b_1} = V_{a_2 a_2 b_2 b_2},
\end{align}
while the exchange integrals between VBS are zero, so the $D_{kl}$ couplings between the reference and the single transfer give no contribution. There are non-zero $P_{kl}$ elements between VBS, in particular there are the expected \emph{forward scattering}
\begin{align}
	- P_{a_1 b_1} = P_{a_1 b_2} = P_{a_2 b_1} = - P_{b_2 a_2} = \frac{1}{2},
\end{align}
in addition to much smaller \emph{backward scattering}
\begin{align}
	- P_{b_1 a_1} = P_{b_1 a_2} = P_{b_2 a_1} = - P_{b_2 a_2}
\end{align}
elements. However, as these will be weighted by exchange (real pair-transfer) integrals
\begin{align}
	V_{a_1 b_1 b_1 a_1} = V_{a_1 b_2 b_2 a_1} = V_{a_2 b_1 b_1 a_2} = V_{a_2 b_2 b_2 a_2} = 0
\end{align}
when $r$ is large these give no contribution either and hence $\braket{M | \hat{H}_C | M^b_a}=0$. 

For couplings of the reference $\ket{M}$ with doubles, the matrix $J$ always has a second small singular value, though again it is generally much larger than the fundamental $\sigma_N$. As a result, none of the doubles couple to the reference through 1-body elements $\gamma_k$. At large $r$, the coupling between the reference $\ket{M}$ and a double swap $\ket{M^{ab}_{ab}}$ has non-zero elements
\begin{align}
	D_{a_1 b_1} = - D_{a_1 b_2} = - D_{a_2 b_1} = D_{a_2 b_2} = D_{b_1 a_1} = - D_{b_1 a_2} = - D_{b_2 a_1} = D_{b_2 a_2} = \frac{1}{4}
\end{align}
which give zero contribution as the corresponding integrals are again symmetric. There are also very small
\begin{align}
	P_{a_1 b_1} = - P_{a_1 b_2} = - P_{a_2 b_1} = P_{a_2 b_2} = - P_{b_1 a_1} = P_{b_1 a_2} = P_{b_2 a_1} = - P_{b_2 a_2}
\end{align}
elements between VBS, which also give zero contribution as the exchange integrals between VBS go to zero for large $r$. Couplings of the reference $\ket{M}$ with pair plus transfers $\ket{M^{ac}_{ab}}$ and double transfers $\ket{M^{cd}_{ab}}$ only have non-zero contributions from $P_{kl}$ type elements, though the largest of these is on the order of $10^{-6}$. Thus the reference does not couple at all with doubles at large $r$.

For H$_8$, there are a limited set of triple excitations, as well as a quadruple swap. The triples \emph{always} have at least 3 small singular values, with an additional small singular value developed for those involving a transfer. The quadruple swap always has 4 small singular values. As a result, the first and second cofactors of $J$ are always near-zero and there is no coupling to the reference at any value of $r$. 

Following these observations, analogues of the Slater-Condon rules for the RG reference $\ket{M}$ may be stated:
\begin{enumerate}
	\item For a $k$-fold excitation, $J$ \emph{always} has $k$ small singular values. Additional small singular values appear for \emph{each} transfer once $\Delta \varepsilon_a < |g|$: a single transfer acquires one extra small singular value, double transfers acquire two extra small singular values, etc.
	\item If $J$ has one small singular value (in particular $\sigma_N$), there are one- and two-body couplings.
	\item If $J$ has two small singular values, only two-body couplings are present.
	\item If $J$ has more than two small singular values, there is no coupling for a two-body operator.
\end{enumerate}

\section{Perturbation theory}
Having established that a given RG reference $\ket{M}$ has no couplings beyond doubles, and thus that RGCISD $\approx$ DOCI, it would be nice to reduce the calculation to something actually feasible. Building the RGCISD matrix means computing the complete TDM ($\mathcal{O}(N^4)$) for the $M(N-M)$ single excitations, and $\binom{M}{2}\binom{N-M}{2}$ double excitations, which scales like $\mathcal{O}(N^4 M^8)$. Even building the RGCIS matrix has a cost on the order of $\mathcal{O}(N^4 M^4)$, which is more expensive than the resulting matrix diagonalization. As the RGCISD wavefunction is dominated by the reference $\ket{M}$, single reference perturbation theory (PT) would seem to be an attractive alternative. 

The standard Rayleigh-Schr\"{o}dinger (RS) PT construction decomposes the Hamiltonian we wish to solve, in this case $\hat{H}_C$ the Coulomb Hamiltonian \eqref{eq:C_ham}, as an exactly solvable reference $\hat{H}_0$ plus a perturbation
\begin{align}
	\hat{H}_C = \hat{H}_0 + \lambda (\hat{H}_C - \hat{H}_0)
\end{align}
and builds energetic corrections order by order with the eigenvectors of $\hat{H}_0$
\begin{align}
	\hat{H}_0 \ket{\psi^{(0)}_{\alpha}} = E^{(0)}_{\alpha} \ket{\psi^{(0)}_{\alpha}}.
\end{align}
From a given reference $\ket{\psi^{(0)}_{0}}$, the 2nd order correction to the energy is
\begin{align} \label{eq:rspt2}
	E^{(2)}_{RS} = \sum_{\alpha \neq 0} \frac{| \braket{\psi^{(0)}_{\alpha} | \hat{H}_C | \psi^{(0)}_{0}} |^2}{E^{(0)}_0 - E^{(0)}_{\alpha}}.
\end{align}
In the present case, the chosen reference would be $\hat{H}_0 = \hat{H}_{VB}$ \eqref{eq:h_vb} which is \emph{far} from $\hat{H}_C$. Worse, as $\ket{\psi^{(0)}_{0}} = \ket{M}$ is not the ground state of \eqref{eq:h_vb}, the 2nd order correction \eqref{eq:rspt2} can be positive. For H$_8$ this does happen, and is so disastrous the curve won't be presented. Such failure was also observed for RSPT corrections to RG states for individual electrons.\cite{carrier:2020,moisset:2022b}

A much better treatment is obtained by defining $\hat{H}_0$ in terms of the expected values of the target $\hat{H}_C$
\begin{align}
	\hat{H}_0 = \sum_{\alpha} \ket{\psi^{(0)}_{\alpha}} \braket{\psi^{(0)}_{\alpha} | \hat{H}_C | \psi^{(0)}_{\alpha}} \bra{\psi^{(0)}_{\alpha}} 
	          = \sum_{\alpha} E_{\alpha} \ket{\psi^{(0)}_{\alpha}}\bra{\psi^{(0)}_{\alpha}}, 
\end{align}
which is the Epstein\cite{epstein:1926}-Nesbet\cite{nesbet:1955} (EN) partitioning. The state $\ket{M}$ is the now the ground state of $\hat{H}_0$, and the 2nd order energy correction is
\begin{align}
	E^{(2)}_{EN} = \sum_{\alpha \neq 0} \frac{| \braket{\psi^{(0)}_{\alpha} | \hat{H}_C | \psi^{(0)}_{0}} |^2}{E_0 - E_{\alpha}}.
\end{align}
As only TDM elements involving the reference are required, this correction is computable with $\mathcal{O}(N^4 M^2)$ cost. ENPT is known to have issues with size-consistency. In the present case, the results appear to be size-consistent so long as the orbitals are localized. If the orbitals are not localized, then RG states are not size-consitent either.
 
\begin{figure} [ht!]
	\begin{subfigure}{\textwidth}
		\centering
		\includegraphics[width=0.49\textwidth]{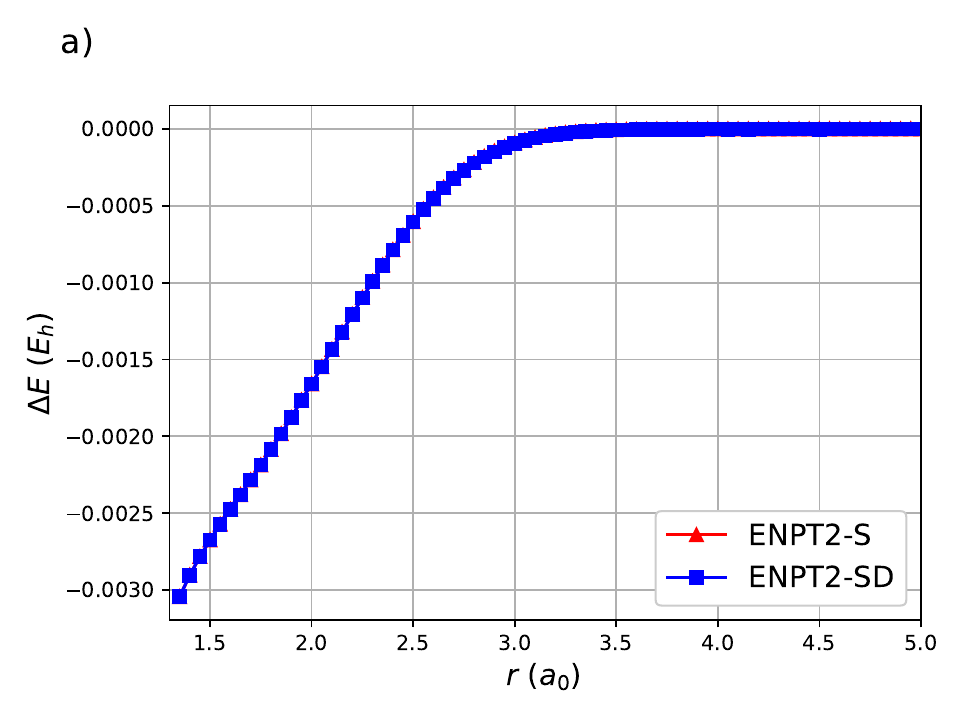} \hfill
		\includegraphics[width=0.49\textwidth]{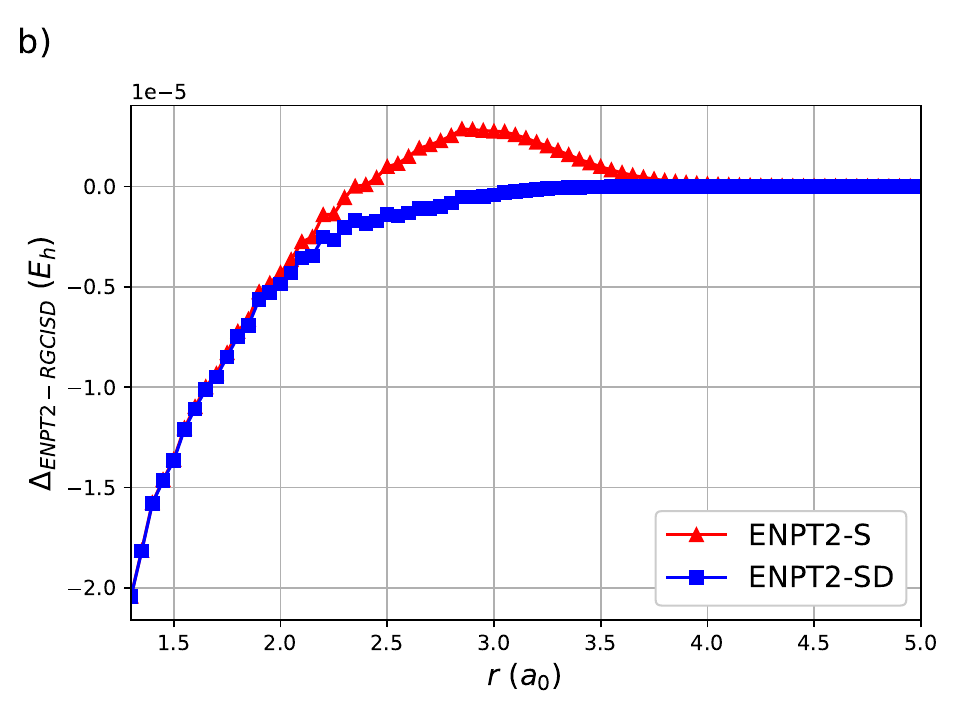}
	\end{subfigure}
	\caption{ENPT2 corrections for linear H$_8$. (a) Perturbative ENPT2 corrections computed with singles, and singles and doubles. (b) Difference of ENPT2 corrections with RGCISD. Results computed with the OO-DOCI orbitals in the STO-6G basis set.}
	\label{fig:h8_pt}
\end{figure}
ENPT2 corrections employing single and double excitations for H$_8$ are shown in Figure \eqref{fig:h8_pt}. Raw energies are not plotted as it is impossible to discern the curves. Notice that the difference between ENPT2 using singles is very close to RGCISD. 

\begin{figure} [ht!]
	\begin{subfigure}{\textwidth}
		\centering
		\includegraphics[width=0.49\textwidth]{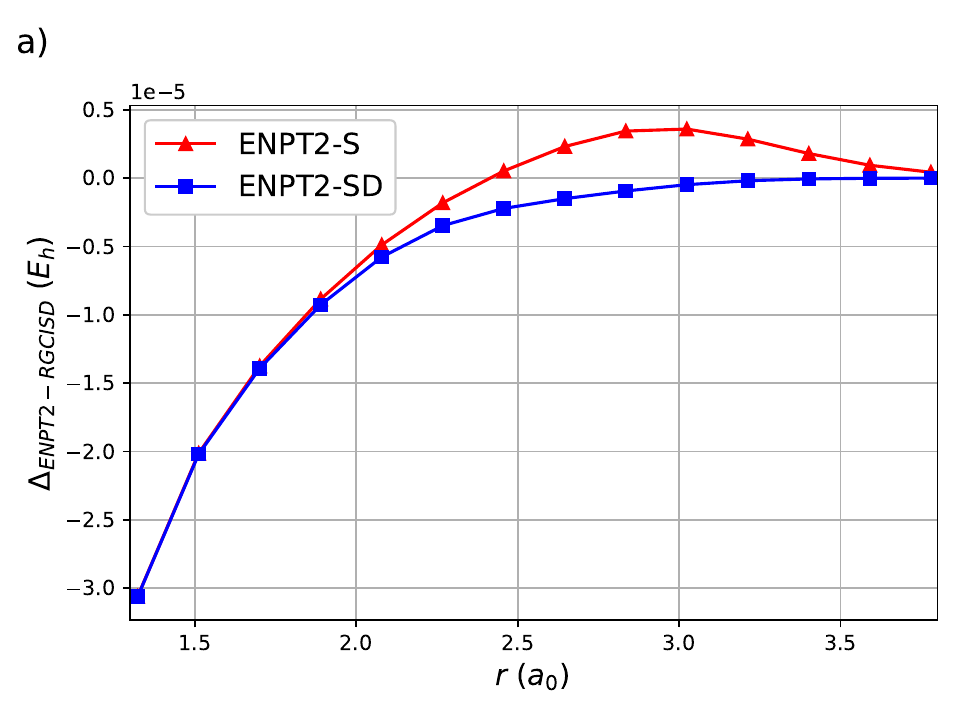} \hfill
		\includegraphics[width=0.49\textwidth]{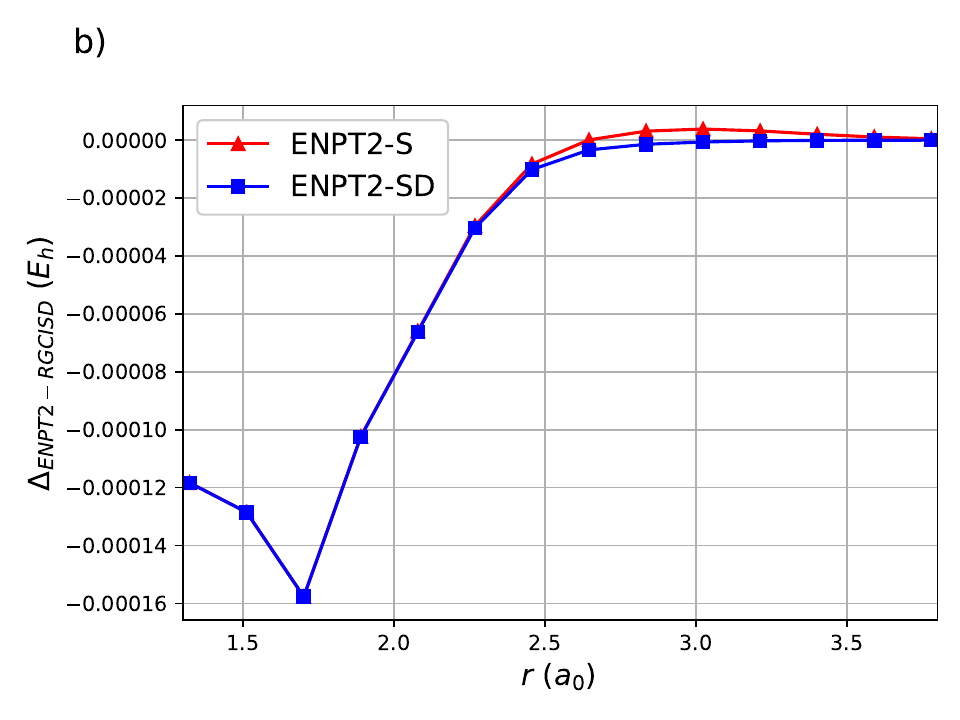} \\
		\includegraphics[width=0.49\textwidth]{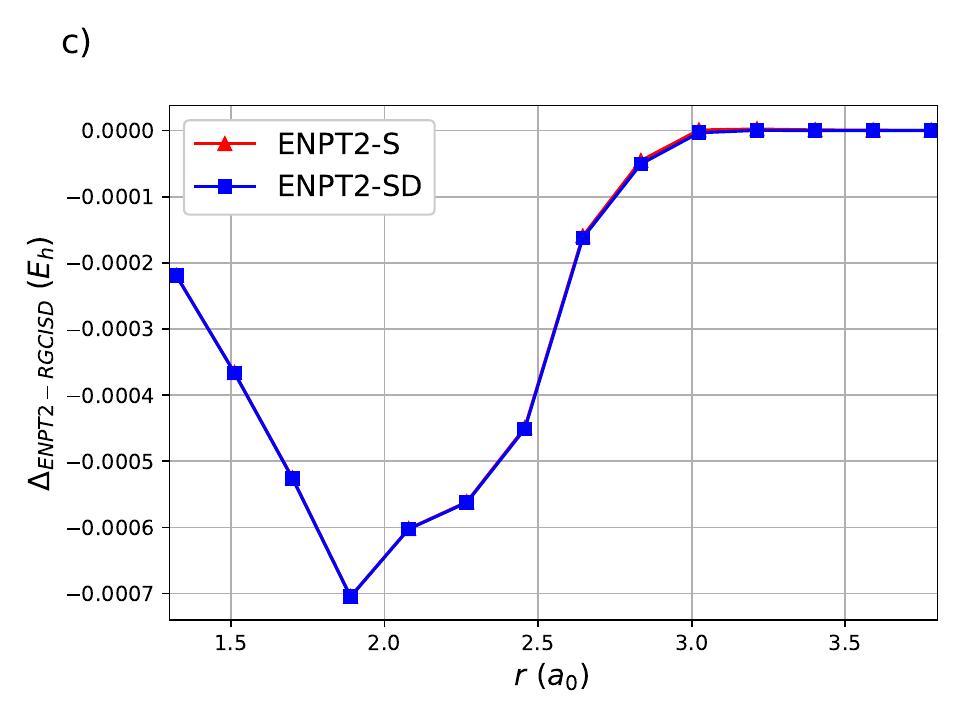} \hfill
		\includegraphics[width=0.49\textwidth]{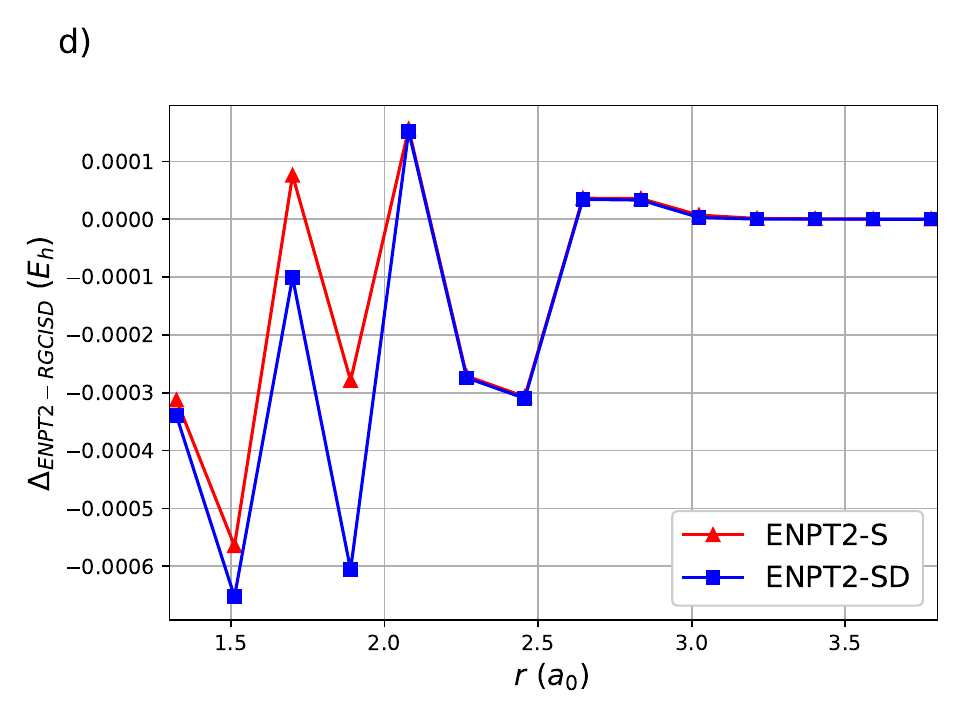}
	\end{subfigure}
	\caption{Difference between ENPT2 and RGCISD for Stair-Evangelista isomers of H$_{10}$: (a) 1D chain, (b) 1D ring, (c) 2D sheet, (d) 3D pyramid. Results computed with the OO-DOCI orbitals in the STO-6G basis set.}
	\label{fig:h10_pt}
\end{figure}
The same is generally true for the Stair-Evangelista isomers of H$_{10}$\cite{stair:2020}: the \emph{chain} of 10 equidistant H atoms, the circular \emph{ring} of 10 equidistant H atoms, the 3-4-3 planar \emph{sheet} of 10 equidistant H atoms and the \emph{pyramid} of 10 equidistant H atoms. The pyramid is in fact a tetrahedron with an H atom at each vertex and an H atom at the midpoint of each edge. OO-DOCI is a reasonable description for the chain and the ring, but not for the sheet nor the pyramid. In all cases, RGCISD $\approx$ DOCI, with an agreement on the order of $10^{-9}$ for the chain and the ring, and $10^{-6}$ for the sheet and the pyramid.\cite{johnson:2023} Figure 5 shows the errors of ENPT2 with respect to RGCISD. In all cases the disagreement between the two is less than 1 mE$_h$. Note that for the first three points of the sheet, and the first four points of the pyramid, there is a curve crossing and the refence RG state is different.\cite{johnson:2023} Even so, the agreement between ENPT2 with singles and RGCISD is quite good for a fraction of the cost.

\begin{figure} [ht!]
	\begin{subfigure}{\textwidth}
		\centering
		\includegraphics[width=0.49\textwidth]{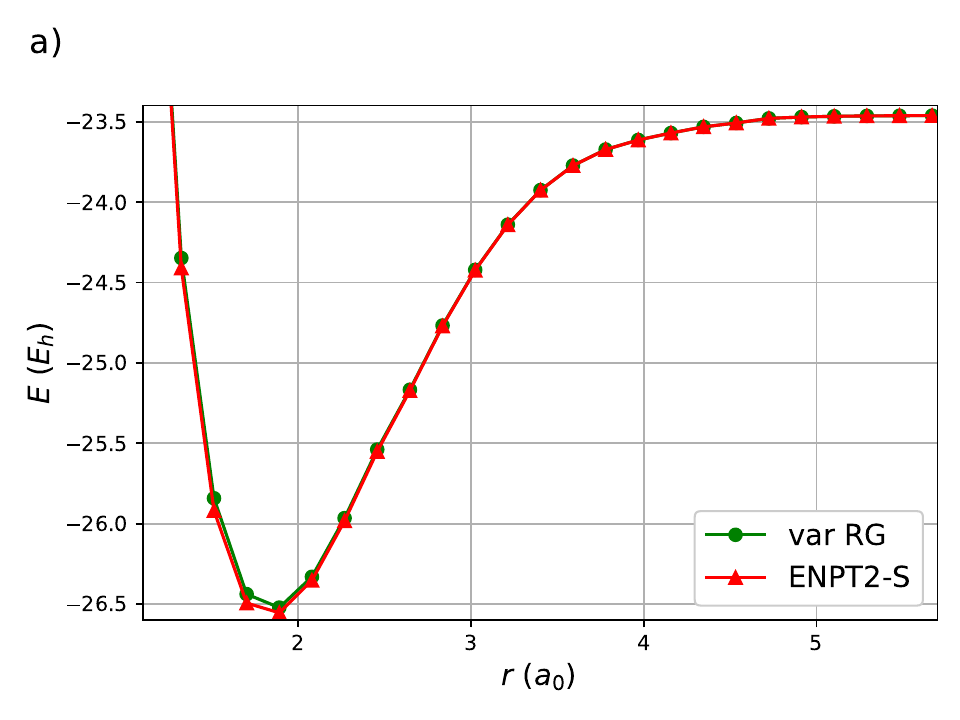} \hfill
		\includegraphics[width=0.49\textwidth]{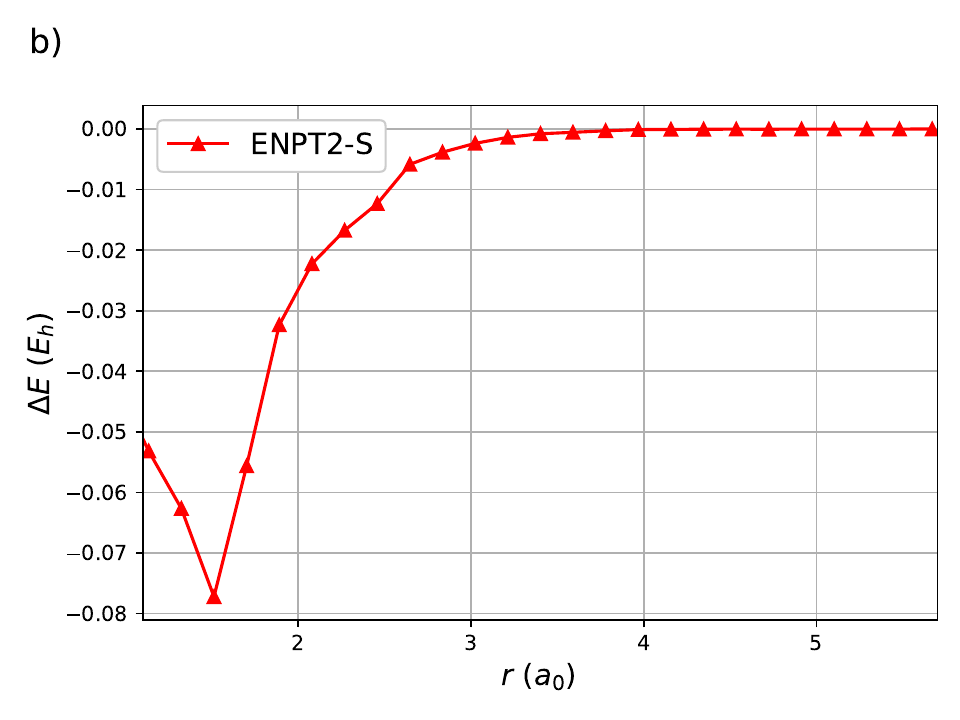}
	\end{subfigure}
	\caption{Variational RG $\ket{M}$ and ENPT2 with singles for linear equidistant H$_{50}$. Results computed with the PNOF7 orbitals in the STO-6G basis set.}
	\label{fig:h50_energy}
\end{figure}
Finally, a variational RG treatment and ENPT2 with singles of linear equidistant H$_{50}$ is shown in figure \ref{fig:h50_energy}. This system is too large to treat with DOCI so the main point is that ENPT2 can be computed in a reasonable time. By computing the TDM elements in parallel, this computation requires roughly an hour on a modest desktop machine. As expected, the reference RG state is $\ket{M}$, and the ENPT2 correction is largest near the equilibrium geometry before decaying to zero at dissociation.  As OO-DOCI is unfeasible, the orbitals employed are those optimal for the Piris natural orbital functional PNOF7\cite{piris:2017,mitxelena:2020a,mitxelena:2020b} from ref.\cite{mitxelena:2022} Natural orbital functionals are close cousins of geminal wavefunctions: the 2-RDM elements are explicit functions of the 1-RDM elements, which is the case for the antisymmetrized product of strongly orthogonal geminals (APSG)\cite{hurley:1953,kutzelnigg:1964,nicely:1971} and the antisymmetrized geminal power (AGP).\cite{coleman:1965,coleman:1989,coleman:1997} PNOF7 is not N-representable, but can be loosely understood as having intrapair elements like APSG and interpair elements like AGP.\cite{moisset:2022a} 

\begin{table}[ht!]
	\centering
	\begin{tabular}{c|c|c|c|c|c}
		$a$ & $b$ & $\sigma_a$ & $\sigma_b$ & $R^2$ & $r_C$   \\
		\hline
		-0.986 & 3.029 & 0.003 & 0.011 & 0.9999 & 3.074 \\
		-0.984 & 3.026 & 0.002 & 0.009 & 0.9999 & 3.074 \\
		-1.080 & 3.504 & 0.008 & 0.033 & 0.9990 & 3.245 \\
		-1.078 & 3.501 & 0.008 & 0.032 & 0.9990 & 3.246 \\
		-1.086 & 3.536 & 0.009 & 0.037 & 0.9987 & 3.255 \\
		-1.083 & 3.527 & 0.008 & 0.033 & 0.9990 & 3.255 \\
		-1.084 & 3.529 & 0.008 & 0.033 & 0.9990 & 3.255 \\
		-1.084 & 3.530 & 0.008 & 0.032 & 0.9990 & 3.255 \\
		-1.084 & 3.530 & 0.008 & 0.032 & 0.9991 & 3.255 \\
		-1.086 & 3.526 & 0.009 & 0.035 & 0.9989 & 3.256 \\ 
		-1.085 & 3.533 & 0.008 & 0.034 & 0.9989 & 3.257 \\
		-1.083 & 3.257 & 0.009 & 0.034 & 0.9989 & 3.257 \\
		-1.064 & 3.475 & 0.008 & 0.034 & 0.9989 & 3.267 \\
		-1.085 & 3.578 & 0.017 & 0.070 & 0.9954 & 3.299 \\
		-1.071 & 3.534 & 0.020 & 0.080 & 0.9939 & 3.300 \\
		-1.074 & 3.547 & 0.018 & 0.074 & 0.9948 & 3.302 \\
		-1.074 & 3.547 & 0.019 & 0.076 & 0.9945 & 3.304 \\
		-1.069 & 3.534 & 0.019 & 0.077 & 0.9942 & 3.307 \\
		-1.070 & 3.540 & 0.019 & 0.077 & 0.9943 & 3.308 \\
		-1.069 & 3.536 & 0.019 & 0.077 & 0.9942 & 3.308 \\
		-1.068 & 3.534 & 0.019 & 0.078 & 0.9941 & 3.308 \\
		-1.069 & 3.538 & 0.019 & 0.077 & 0.9943 & 3.308 \\
		-1.070 & 3.541 & 0.019 & 0.078 & 0.9941 & 3.309 \\
		-1.071 & 3.561 & 0.017 & 0.067 & 0.9956 & 3.326 \\
		-1.076 & 3.580 & 0.016 & 0.065 & 0.9959 & 3.327
	\end{tabular} 
	\caption{\label{table:regression} Linear regression of $\ln \Delta \varepsilon$ in units of $\vert g \vert$ as a function of $r$: $\ln \frac{ \Delta \varepsilon}{\vert g \vert} = a r + b$. Standard errors of the slope ($\sigma_a$), and y-intercept ($\sigma_b$) as well as the correlation coefficient ($R^2$) are reported. The x-intercept is computed as $r_0 = -\frac{b}{a}$.}
\end{table}
Since the RG reference is $\ket{M}$, a linear regression of $\ln \frac{ \Delta \varepsilon}{\vert g \vert}$ may be performed, which is summarized in Table \ref{table:regression}. The PNOF7 orbitals change their nature at small $r$, so the regression may only be performed in the linear regime, which occurs once $r>2.08$ bohr. The fit of the parameters is quite good, though many of correlation coefficients ($R^2$) are not perfect. All of the pairs have $r_C$ between $3.07$ bohr and $3.23$ bohr. Again, for the moment the purpose of the H$_{50}$ results is to demonstrate that they are feasible. While PNOF7 orbitals are likely a reasonable approximation, a proper orbital optimization will be included for the RG variational treatment in the near future.

\section{Conclusion}
Tractable expressions for the TDM elements between RG states have been made possible by isolating the near-zero singular values in the inverse of $J$, a trick developed in the robust Wick's theorem formalism of Chen and Scuseria.\cite{chen:2023} These TDM elements may be computed perfectly in parallel. Couplings of the RG reference $\ket{M}$ with its low-lying excited states all vanish for large $r$ in different ways. As expected, localized orbitals are necessary for size-consistency. The near-zero singular values of the effective overlap matrix $J$ and the nature of the RG excited states provide analogues of the Slater-Condon rules for Slater determinants. In particular, a $k$-pair excitation always has $k$ near-zero singular values, with additional small singular value for each transfer between VBS when $\Delta \varepsilon_a < |g|$. If $J$ has three or more small singular values, there is no coupling. Finally, Epstein-Nesbet perturbation theory with pair-single excitations yields results comparable to RGCISD at a much reduced cost. While all of these results are presented for seniority-zero states, many of them will carry forward to RG excited states with different seniorities which will be treated in an upcoming contribution.
\begin{acknowledgement}
The author thanks Mario Piris for sharing the PNOF7 orbitals for H$_{50}$, and Guo Chen for many discussions concerning their robust formulation of Wick's theorem.\cite{chen:2023} This research was supported by NSERC and the Digital Research Alliance of Canada. 
\end{acknowledgement}



\bibliography{gusvd}

\providecommand{\latin}[1]{#1}
\makeatletter
\providecommand{\doi}
  {\begingroup\let\do\@makeother\dospecials
  \catcode`\{=1 \catcode`\}=2 \doi@aux}
\providecommand{\doi@aux}[1]{\endgroup\texttt{#1}}
\makeatother
\providecommand*\mcitethebibliography{\thebibliography}
\csname @ifundefined\endcsname{endmcitethebibliography}
  {\let\endmcitethebibliography\endthebibliography}{}
\begin{mcitethebibliography}{108}
\providecommand*\natexlab[1]{#1}
\providecommand*\mciteSetBstSublistMode[1]{}
\providecommand*\mciteSetBstMaxWidthForm[2]{}
\providecommand*\mciteBstWouldAddEndPuncttrue
  {\def\EndOfBibitem{\unskip.}}
\providecommand*\mciteBstWouldAddEndPunctfalse
  {\let\EndOfBibitem\relax}
\providecommand*\mciteSetBstMidEndSepPunct[3]{}
\providecommand*\mciteSetBstSublistLabelBeginEnd[3]{}
\providecommand*\EndOfBibitem{}
\mciteSetBstSublistMode{f}
\mciteSetBstMaxWidthForm{subitem}{(\alph{mcitesubitemcount})}
\mciteSetBstSublistLabelBeginEnd
  {\mcitemaxwidthsubitemform\space}
  {\relax}
  {\relax}

\bibitem[\v{C}\'{i}\v{z}ek(1966)]{cizek:1966}
\v{C}\'{i}\v{z}ek,~J. On the Correlation Problem in Atomic and Molecular
  Systems. Calculation of Wavefunction Components in Ursell-Type Expansion
  Using Quantum-Field Theoretical Methods. \emph{J. Chem. Phys.} \textbf{1966},
  \emph{45}, 4256--4266\relax
\mciteBstWouldAddEndPuncttrue
\mciteSetBstMidEndSepPunct{\mcitedefaultmidpunct}
{\mcitedefaultendpunct}{\mcitedefaultseppunct}\relax
\EndOfBibitem
\bibitem[\v{C}\'{i}\v{z}ek and Paldus(1971)\v{C}\'{i}\v{z}ek, and
  Paldus]{cizek:1971}
\v{C}\'{i}\v{z}ek,~J.; Paldus,~J. Correlation Problems in Atomic and Molecular
  Systems III. Rederivation of the Coupled-Pair Many-Electron Theory Using the
  Traditional Quantum Chemical Methods. \emph{Int. J. Quantum Chem.}
  \textbf{1971}, \emph{5}, 359--379\relax
\mciteBstWouldAddEndPuncttrue
\mciteSetBstMidEndSepPunct{\mcitedefaultmidpunct}
{\mcitedefaultendpunct}{\mcitedefaultseppunct}\relax
\EndOfBibitem
\bibitem[Purvis~III and Bartlett(1982)Purvis~III, and Bartlett]{purvis_1982}
Purvis~III,~G.~D.; Bartlett,~R.~J. A Full Coupled-Cluster Singles and Doubles
  Model: The Inclusion of Disconnected Triples. \emph{J. Chem. Phys.}
  \textbf{1982}, \emph{76}, 1910--1918\relax
\mciteBstWouldAddEndPuncttrue
\mciteSetBstMidEndSepPunct{\mcitedefaultmidpunct}
{\mcitedefaultendpunct}{\mcitedefaultseppunct}\relax
\EndOfBibitem
\bibitem[Shavitt and Bartlett(2009)Shavitt, and Bartlett]{bartlett_book}
Shavitt,~I.; Bartlett,~R.~J. \emph{Many-Body Methods in Chemistry and Physics};
  Cambridge University Press: Cambridge, 2009\relax
\mciteBstWouldAddEndPuncttrue
\mciteSetBstMidEndSepPunct{\mcitedefaultmidpunct}
{\mcitedefaultendpunct}{\mcitedefaultseppunct}\relax
\EndOfBibitem
\bibitem[Bartlett and Musia{\l}(2007)Bartlett, and Musia{\l}]{bartlett:2007}
Bartlett,~R.~J.; Musia{\l},~M. Coupled-Cluster Theory in Quantum Chemistry.
  \emph{Rev. Mod. Phys.} \textbf{2007}, \emph{79}, 291--352\relax
\mciteBstWouldAddEndPuncttrue
\mciteSetBstMidEndSepPunct{\mcitedefaultmidpunct}
{\mcitedefaultendpunct}{\mcitedefaultseppunct}\relax
\EndOfBibitem
\bibitem[Helgaker \latin{et~al.}(2000)Helgaker, J{\o}rgenson, and
  Olsen]{helgaker_book}
Helgaker,~T.; J{\o}rgenson,~P.; Olsen,~J. \emph{Molecular Electronic-Structure
  Theory}; Wiley \& Sons: West Sussex, 2000\relax
\mciteBstWouldAddEndPuncttrue
\mciteSetBstMidEndSepPunct{\mcitedefaultmidpunct}
{\mcitedefaultendpunct}{\mcitedefaultseppunct}\relax
\EndOfBibitem
\bibitem[Roos \latin{et~al.}(1980)Roos, Taylor, and Siegbahn]{roos:1980}
Roos,~B.~O.; Taylor,~P.~R.; Siegbahn,~P. E.~M. A Complete Active Space SCF
  Method (CASSCF) Using a Density Matrix Formulated Super-CI Approach.
  \emph{Chem. Phys.} \textbf{1980}, \emph{48}, 157--173\relax
\mciteBstWouldAddEndPuncttrue
\mciteSetBstMidEndSepPunct{\mcitedefaultmidpunct}
{\mcitedefaultendpunct}{\mcitedefaultseppunct}\relax
\EndOfBibitem
\bibitem[Siegbahn \latin{et~al.}(1980)Siegbahn, Heiberg, Roos, and
  Levy]{siegbahn:1980}
Siegbahn,~P. E.~M.; Heiberg,~A.; Roos,~B.; Levy,~B. A Comparison of the
  Super-CI and the Newton-Raphson Scheme in the Complete Active Space SCF
  Method. \emph{Phys. Scr.} \textbf{1980}, \emph{21}, 323--327\relax
\mciteBstWouldAddEndPuncttrue
\mciteSetBstMidEndSepPunct{\mcitedefaultmidpunct}
{\mcitedefaultendpunct}{\mcitedefaultseppunct}\relax
\EndOfBibitem
\bibitem[Siegbahn \latin{et~al.}(1981)Siegbahn, Alml\"{o}f, Heiberg, and
  Roos]{siegbahn:1981}
Siegbahn,~P. E.~M.; Alml\"{o}f,~J.; Heiberg,~A.; Roos,~B. The Complete Active
  Space SCF (CASSCF) Method in a Newton-Raphson Formulation with Application to
  the HNO Molecule. \emph{J. Chem. Phys.} \textbf{1981}, \emph{74},
  2384--2396\relax
\mciteBstWouldAddEndPuncttrue
\mciteSetBstMidEndSepPunct{\mcitedefaultmidpunct}
{\mcitedefaultendpunct}{\mcitedefaultseppunct}\relax
\EndOfBibitem
\bibitem[Roos(1987)]{roos:1987}
Roos,~B.~O. \emph{Advances in Chemical Physics}; John Wiley \& Sons, Ltd, 1987;
  pp 399--445\relax
\mciteBstWouldAddEndPuncttrue
\mciteSetBstMidEndSepPunct{\mcitedefaultmidpunct}
{\mcitedefaultendpunct}{\mcitedefaultseppunct}\relax
\EndOfBibitem
\bibitem[Olsen \latin{et~al.}(1988)Olsen, Roos, J{\o}rgenson, and
  Jensen]{olsen:1988}
Olsen,~J.; Roos,~B.~J.; J{\o}rgenson,~P.; Jensen,~H. J.~A. Determinant Based
  Configuration Interaction Algorithms for Complete and Restricted
  Configuration Interaction Spaces. \emph{J. Chem. Phys.} \textbf{1988},
  \emph{89}, 2185--2192\relax
\mciteBstWouldAddEndPuncttrue
\mciteSetBstMidEndSepPunct{\mcitedefaultmidpunct}
{\mcitedefaultendpunct}{\mcitedefaultseppunct}\relax
\EndOfBibitem
\bibitem[Malmqvist \latin{et~al.}(1990)Malmqvist, Rendell, and
  Roos]{malmqvist:1990}
Malmqvist,~P.~A.; Rendell,~A.; Roos,~B.~O. The Restricted Active Space
  Self-Consistent-Field Method, Implemented with a Split Graph Unitary Group
  Approach. \emph{J. Phys. Chem.} \textbf{1990}, \emph{94}, 5477--5482\relax
\mciteBstWouldAddEndPuncttrue
\mciteSetBstMidEndSepPunct{\mcitedefaultmidpunct}
{\mcitedefaultendpunct}{\mcitedefaultseppunct}\relax
\EndOfBibitem
\bibitem[Fleig \latin{et~al.}(2001)Fleig, Olsen, and Marian]{fleig:2001}
Fleig,~T.; Olsen,~J.; Marian,~C.~M. The Generalized Active Space Concept for
  the Relativistic Treatment of Electron Correlation. I. Kramers-Restricted
  Two-Component Configuration Interaction. \emph{J. Chem. Phys.} \textbf{2001},
  \emph{114}, 4775--4790\relax
\mciteBstWouldAddEndPuncttrue
\mciteSetBstMidEndSepPunct{\mcitedefaultmidpunct}
{\mcitedefaultendpunct}{\mcitedefaultseppunct}\relax
\EndOfBibitem
\bibitem[Ma \latin{et~al.}(2011)Ma, Manni, and Gagliardi]{ma:2011}
Ma,~D.; Manni,~G.~L.; Gagliardi,~L. The Generalized Active Space Concept in
  Multiconfigurational Self-Consistent Field Methods. \emph{J. Chem. Phys.}
  \textbf{2011}, \emph{135}, 044128\relax
\mciteBstWouldAddEndPuncttrue
\mciteSetBstMidEndSepPunct{\mcitedefaultmidpunct}
{\mcitedefaultendpunct}{\mcitedefaultseppunct}\relax
\EndOfBibitem
\bibitem[Manni \latin{et~al.}(2013)Manni, Ma, Aquilante, Olsen, and
  Gagliardi]{manni:2013}
Manni,~G.~L.; Ma,~D.; Aquilante,~F.; Olsen,~J.; Gagliardi,~L. SplitGAS Method
  for Strong Correlation and the Challending Case of Cr$_2$. \emph{J. Chem.
  Theory Comput.} \textbf{2013}, \emph{9}, 3375--3384\relax
\mciteBstWouldAddEndPuncttrue
\mciteSetBstMidEndSepPunct{\mcitedefaultmidpunct}
{\mcitedefaultendpunct}{\mcitedefaultseppunct}\relax
\EndOfBibitem
\bibitem[Thomas \latin{et~al.}(2015)Thomas, Sun, Alavi, and Booth]{thomas:2015}
Thomas,~R.~E.; Sun,~Q.; Alavi,~A.; Booth,~G.~H. Stochastic Multiconfigurational
  Self-Consistent Field Theory. \emph{J. Chem. Theory Comput.} \textbf{2015},
  \emph{11}, 5316--5325\relax
\mciteBstWouldAddEndPuncttrue
\mciteSetBstMidEndSepPunct{\mcitedefaultmidpunct}
{\mcitedefaultendpunct}{\mcitedefaultseppunct}\relax
\EndOfBibitem
\bibitem[Manni \latin{et~al.}(2016)Manni, Smart, and Alavi]{manni:2016}
Manni,~G.~L.; Smart,~S.~D.; Alavi,~A. Combining the Complete Active Space
  Self-Consistent Field Method and the Full Configuration Interaction Quantum
  Monte Carlo within a Super-CI Framework, with Application to Challenging
  Metal-Porphyrins. \emph{J. Chem. Theory Comput.} \textbf{2016}, \emph{12},
  1245--1258\relax
\mciteBstWouldAddEndPuncttrue
\mciteSetBstMidEndSepPunct{\mcitedefaultmidpunct}
{\mcitedefaultendpunct}{\mcitedefaultseppunct}\relax
\EndOfBibitem
\bibitem[Schriber and Evangelista(2016)Schriber, and
  Evangelista]{schriber:2016}
Schriber,~J.~B.; Evangelista,~F.~A. Communication: An Adaptive Configuration
  Interaction Approach for Strongly Correlated Electrons with Tunable Accuracy.
  \emph{J. Chem. Phys.} \textbf{2016}, \emph{144}, 161106\relax
\mciteBstWouldAddEndPuncttrue
\mciteSetBstMidEndSepPunct{\mcitedefaultmidpunct}
{\mcitedefaultendpunct}{\mcitedefaultseppunct}\relax
\EndOfBibitem
\bibitem[Levine \latin{et~al.}(2020)Levine, Hait, Tubman, Lehtola, Whaley, and
  Head-Gordon]{levine:2020}
Levine,~D.~S.; Hait,~D.; Tubman,~N.~M.; Lehtola,~S.; Whaley,~K.~B.;
  Head-Gordon,~M. CASSCF with Extremely Large Active Spaces Using the Adaptive
  Sampling Configuration Interaction Method. \emph{J. Chem. Theory Comput.}
  \textbf{2020}, \emph{16}, 2340--2354\relax
\mciteBstWouldAddEndPuncttrue
\mciteSetBstMidEndSepPunct{\mcitedefaultmidpunct}
{\mcitedefaultendpunct}{\mcitedefaultseppunct}\relax
\EndOfBibitem
\bibitem[Chan and Head-Gordon(2002)Chan, and Head-Gordon]{chan:2002}
Chan,~G. K.-L.; Head-Gordon,~M. Highly Correlated Calculations with a
  Polynomial Cost Algorithm: A Study of the Density Matrix Renormalization
  Group. \emph{J. Chem. Phys.} \textbf{2002}, \emph{116}, 4462--4476\relax
\mciteBstWouldAddEndPuncttrue
\mciteSetBstMidEndSepPunct{\mcitedefaultmidpunct}
{\mcitedefaultendpunct}{\mcitedefaultseppunct}\relax
\EndOfBibitem
\bibitem[Ghosh \latin{et~al.}(2008)Ghosh, Hachmann, Yanai, and
  Chan]{ghosh:2008}
Ghosh,~D.; Hachmann,~J.; Yanai,~T.; Chan,~G. K.-L. Orbital Optimization in the
  Density Matrix Renormalization Group, with Applications to Polyenes and
  $\beta$-carotene. \emph{J. Chem. Phys.} \textbf{2008}, \emph{128},
  144117\relax
\mciteBstWouldAddEndPuncttrue
\mciteSetBstMidEndSepPunct{\mcitedefaultmidpunct}
{\mcitedefaultendpunct}{\mcitedefaultseppunct}\relax
\EndOfBibitem
\bibitem[Yanai \latin{et~al.}(2009)Yanai, Kurashige, Ghosh, and
  Chan]{yanai:2009}
Yanai,~T.; Kurashige,~Y.; Ghosh,~D.; Chan,~G. K.-L. Accelerating Convergence in
  Iterative Solution for Large-Scale Complete Active Space
  Self-Consistent-Field Calculations. \emph{Int. J. Quantum Chem.}
  \textbf{2009}, \emph{109}, 2178--2190\relax
\mciteBstWouldAddEndPuncttrue
\mciteSetBstMidEndSepPunct{\mcitedefaultmidpunct}
{\mcitedefaultendpunct}{\mcitedefaultseppunct}\relax
\EndOfBibitem
\bibitem[Wouters \latin{et~al.}(2014)Wouters, Poelmans, Ayers, and
  Van~Neck]{wouters:2014}
Wouters,~S.; Poelmans,~W.; Ayers,~P.~W.; Van~Neck,~D. CheMPS2: A Free
  Open-Source Spin-Adapted Implementation of the Density Matrix Renormalization
  Group for Ab Initio Quantum Chemistry. \emph{Comput. Phys. Commun.}
  \textbf{2014}, \emph{185}, 1501--1514\relax
\mciteBstWouldAddEndPuncttrue
\mciteSetBstMidEndSepPunct{\mcitedefaultmidpunct}
{\mcitedefaultendpunct}{\mcitedefaultseppunct}\relax
\EndOfBibitem
\bibitem[Sun \latin{et~al.}(2017)Sun, Yang, and Chan]{sun:2017}
Sun,~Q.; Yang,~J.; Chan,~G. K.-L. A General Second Order Complete Active Space
  Self-Consistent-Field Solver for Large-Scale Systems. \emph{Chem. Phys.
  Lett.} \textbf{2017}, \emph{683}, 291--299\relax
\mciteBstWouldAddEndPuncttrue
\mciteSetBstMidEndSepPunct{\mcitedefaultmidpunct}
{\mcitedefaultendpunct}{\mcitedefaultseppunct}\relax
\EndOfBibitem
\bibitem[Ma \latin{et~al.}(2017)Ma, Knecht, Keller, and Reiher]{ma:2017}
Ma,~Y.; Knecht,~S.; Keller,~S.; Reiher,~M. Second-Order Self-Consistent-Field
  Density-Matrix Renormalization Group. \emph{J. Chem. Theory Comput.}
  \textbf{2017}, \emph{13}, 2533--2549\relax
\mciteBstWouldAddEndPuncttrue
\mciteSetBstMidEndSepPunct{\mcitedefaultmidpunct}
{\mcitedefaultendpunct}{\mcitedefaultseppunct}\relax
\EndOfBibitem
\bibitem[Fock(1950)]{fock:1950}
Fock,~V. Application of Two-Electron Functions to the Theory of Chemical Bonds.
  \emph{Dokl. Akad. Nauk SSSR} \textbf{1950}, \emph{73}, 735--739\relax
\mciteBstWouldAddEndPuncttrue
\mciteSetBstMidEndSepPunct{\mcitedefaultmidpunct}
{\mcitedefaultendpunct}{\mcitedefaultseppunct}\relax
\EndOfBibitem
\bibitem[Silver(1969)]{silver:1969}
Silver,~D.~M. Natural Orbital Expansion of Interacting Geminals. \emph{J. Chem.
  Phys.} \textbf{1969}, \emph{50}, 5108--5116\relax
\mciteBstWouldAddEndPuncttrue
\mciteSetBstMidEndSepPunct{\mcitedefaultmidpunct}
{\mcitedefaultendpunct}{\mcitedefaultseppunct}\relax
\EndOfBibitem
\bibitem[Silver(1970)]{silver:1970a}
Silver,~D.~M. Bilinear Orbital Expansion of Geminal-Product Correlated
  Wavefunctions. \emph{J. Chem. Phys.} \textbf{1970}, \emph{52}, 299--303\relax
\mciteBstWouldAddEndPuncttrue
\mciteSetBstMidEndSepPunct{\mcitedefaultmidpunct}
{\mcitedefaultendpunct}{\mcitedefaultseppunct}\relax
\EndOfBibitem
\bibitem[Silver \latin{et~al.}(1970)Silver, Mehler, and
  Ruedenberg]{silver:1970b}
Silver,~D.~M.; Mehler,~E.~L.; Ruedenberg,~K. Electron Correlation and Separated
  Pair Approximation in Diatomic Molecules. I. Theory. \emph{J. Chem. Phys.}
  \textbf{1970}, \emph{52}, 1174--1180\relax
\mciteBstWouldAddEndPuncttrue
\mciteSetBstMidEndSepPunct{\mcitedefaultmidpunct}
{\mcitedefaultendpunct}{\mcitedefaultseppunct}\relax
\EndOfBibitem
\bibitem[Silver \latin{et~al.}(1970)Silver, Ruedenberg, and
  Mehler]{silver:1970c}
Silver,~D.~M.; Ruedenberg,~K.; Mehler,~E.~L. Electron Correlation and Separated
  Pair Approximation in Diatomic Molecules. III. Imidogen. \emph{J. Chem.
  Phys.} \textbf{1970}, \emph{52}, 1206--1227\relax
\mciteBstWouldAddEndPuncttrue
\mciteSetBstMidEndSepPunct{\mcitedefaultmidpunct}
{\mcitedefaultendpunct}{\mcitedefaultseppunct}\relax
\EndOfBibitem
\bibitem[Paldus(1972)]{paldus:1972a}
Paldus,~J. Diagrammatic Method for Geminals. I. Theory. \emph{J. Chem. Phys.}
  \textbf{1972}, \emph{57}, 638--651\relax
\mciteBstWouldAddEndPuncttrue
\mciteSetBstMidEndSepPunct{\mcitedefaultmidpunct}
{\mcitedefaultendpunct}{\mcitedefaultseppunct}\relax
\EndOfBibitem
\bibitem[Paldus \latin{et~al.}(1972)Paldus, Sengupta, and
  \v{C}\'{i}\v{z}ek]{paldus:1972b}
Paldus,~J.; Sengupta,~S.; \v{C}\'{i}\v{z}ek,~J. Diagrammatic Method for
  Geminals. II. Applications. \emph{J. Chem. Phys.} \textbf{1972}, \emph{57},
  652--666\relax
\mciteBstWouldAddEndPuncttrue
\mciteSetBstMidEndSepPunct{\mcitedefaultmidpunct}
{\mcitedefaultendpunct}{\mcitedefaultseppunct}\relax
\EndOfBibitem
\bibitem[Moisset \latin{et~al.}(2022)Moisset, Fecteau, and
  Johnson]{moisset:2022a}
Moisset,~J.-D.; Fecteau,~C.-E.; Johnson,~P.~A. Density Matrices of
  Seniority-Zero Geminal Wavefunctions. \emph{J. Chem. Phys.} \textbf{2022},
  \emph{156}, 214110\relax
\mciteBstWouldAddEndPuncttrue
\mciteSetBstMidEndSepPunct{\mcitedefaultmidpunct}
{\mcitedefaultendpunct}{\mcitedefaultseppunct}\relax
\EndOfBibitem
\bibitem[Limacher \latin{et~al.}(2013)Limacher, Ayers, Johnson,
  De~Baerdemacker, Van~Neck, and Bultinck]{limacher:2013}
Limacher,~P.~A.; Ayers,~P.~W.; Johnson,~P.~A.; De~Baerdemacker,~S.;
  Van~Neck,~D.; Bultinck,~P. A New Mean-Field Method Suitable for Strongly
  Correlated Electrons: Computationally Facile Antisymmetric Products of
  Nonorthogonal Geminals. \emph{J. Chem. Theory Comput.} \textbf{2013},
  \emph{9}, 1394--1401\relax
\mciteBstWouldAddEndPuncttrue
\mciteSetBstMidEndSepPunct{\mcitedefaultmidpunct}
{\mcitedefaultendpunct}{\mcitedefaultseppunct}\relax
\EndOfBibitem
\bibitem[Stein \latin{et~al.}(2014)Stein, Henderson, and Scuseria]{stein:2014}
Stein,~T.; Henderson,~T.~M.; Scuseria,~G.~E. Seniority Zero Pair Coupled
  Cluster Doubles Theory. \emph{J. Chem. Phys.} \textbf{2014}, \emph{140},
  214113\relax
\mciteBstWouldAddEndPuncttrue
\mciteSetBstMidEndSepPunct{\mcitedefaultmidpunct}
{\mcitedefaultendpunct}{\mcitedefaultseppunct}\relax
\EndOfBibitem
\bibitem[Limacher \latin{et~al.}(2014)Limacher, Kim, Ayers, Johnson,
  De~Baerdemacker, Van~Neck, and Bultinck]{limacher:2014a}
Limacher,~P.~A.; Kim,~T.~D.; Ayers,~P.~W.; Johnson,~P.~A.; De~Baerdemacker,~S.;
  Van~Neck,~D.; Bultinck,~P. The Influence of Orbital Rotation on the Energy of
  Closed-Shell Wavefunctions. \emph{Mol. Phys.} \textbf{2014}, \emph{112},
  853--862\relax
\mciteBstWouldAddEndPuncttrue
\mciteSetBstMidEndSepPunct{\mcitedefaultmidpunct}
{\mcitedefaultendpunct}{\mcitedefaultseppunct}\relax
\EndOfBibitem
\bibitem[Limacher \latin{et~al.}(2014)Limacher, Ayers, Johnson,
  De~Baerdemacker, Van~Neck, and Bultinck]{limacher:2014b}
Limacher,~P.~A.; Ayers,~P.~W.; Johnson,~P.~A.; De~Baerdemacker,~S.;
  Van~Neck,~D.; Bultinck,~P. Simple and Inexpensive Perturbative Correction
  Schemes for Antisymmetric Products of Nonorthogonal Geminals. \emph{Phys.
  Chem. Chem. Phys.} \textbf{2014}, \emph{16}, 5061--5065\relax
\mciteBstWouldAddEndPuncttrue
\mciteSetBstMidEndSepPunct{\mcitedefaultmidpunct}
{\mcitedefaultendpunct}{\mcitedefaultseppunct}\relax
\EndOfBibitem
\bibitem[Henderson \latin{et~al.}(2014)Henderson, Scuseria, Dukelsky,
  Signoracci, and Duguet]{henderson:2014a}
Henderson,~T.~M.; Scuseria,~G.~E.; Dukelsky,~J.; Signoracci,~A.; Duguet,~T.
  Quasiparticle Coupled Cluster Theory for Pairing Interactions. \emph{Phys.
  Rev. C} \textbf{2014}, \emph{89}, 054305\relax
\mciteBstWouldAddEndPuncttrue
\mciteSetBstMidEndSepPunct{\mcitedefaultmidpunct}
{\mcitedefaultendpunct}{\mcitedefaultseppunct}\relax
\EndOfBibitem
\bibitem[Henderson \latin{et~al.}(2014)Henderson, Bulik, Stein, and
  Scuseria]{henderson:2014b}
Henderson,~T.~M.; Bulik,~I.~W.; Stein,~T.; Scuseria,~G.~E. Seniority-Based
  Coupled Cluster Theory. \emph{J. Chem. Phys.} \textbf{2014}, \emph{141},
  244104\relax
\mciteBstWouldAddEndPuncttrue
\mciteSetBstMidEndSepPunct{\mcitedefaultmidpunct}
{\mcitedefaultendpunct}{\mcitedefaultseppunct}\relax
\EndOfBibitem
\bibitem[Boguslawski \latin{et~al.}(2014)Boguslawski, Tecmer, Ayers, Bultinck,
  De~Baerdemacker, and Van~Neck]{boguslawski:2014a}
Boguslawski,~K.; Tecmer,~P.; Ayers,~P.~W.; Bultinck,~P.; De~Baerdemacker,~S.;
  Van~Neck,~D. Efficient Description of Strongly Correlated Electrons with
  Mean-Field Cost. \emph{Phys. Rev. B} \textbf{2014}, \emph{89},
  201106(R)\relax
\mciteBstWouldAddEndPuncttrue
\mciteSetBstMidEndSepPunct{\mcitedefaultmidpunct}
{\mcitedefaultendpunct}{\mcitedefaultseppunct}\relax
\EndOfBibitem
\bibitem[Boguslawski \latin{et~al.}(2014)Boguslawski, Tecmer, Bultinck,
  De~Baerdemacker, Van~Neck, and Ayers]{boguslawski:2014b}
Boguslawski,~K.; Tecmer,~P.; Bultinck,~P.; De~Baerdemacker,~S.; Van~Neck,~D.;
  Ayers,~P.~W. Nonvariational Orbital Optimization Techniques for the AP1roG
  Wave Function. \emph{J. Chem. Theory Comput.} \textbf{2014}, \emph{10},
  4873--4882\relax
\mciteBstWouldAddEndPuncttrue
\mciteSetBstMidEndSepPunct{\mcitedefaultmidpunct}
{\mcitedefaultendpunct}{\mcitedefaultseppunct}\relax
\EndOfBibitem
\bibitem[Boguslawski \latin{et~al.}(2014)Boguslawski, Tecmer, Limacher,
  Johnson, Ayers, Bultinck, De~Baerdemacker, and Van~Neck]{boguslawski:2014c}
Boguslawski,~K.; Tecmer,~P.; Limacher,~P.~A.; Johnson,~P.~A.; Ayers,~P.~W.;
  Bultinck,~P.; De~Baerdemacker,~S.; Van~Neck,~D. Projected Seniority-Two
  Orbital Optimization of the Antisymmetric Product of One-Reference Orbital
  Geminal. \emph{J. Chem. Phys.} \textbf{2014}, \emph{140}, 214114\relax
\mciteBstWouldAddEndPuncttrue
\mciteSetBstMidEndSepPunct{\mcitedefaultmidpunct}
{\mcitedefaultendpunct}{\mcitedefaultseppunct}\relax
\EndOfBibitem
\bibitem[Tecmer \latin{et~al.}(2014)Tecmer, Boguslawski, Johnson, Limacher,
  Chan, Verstraelen, and Ayers]{tecmer:2014}
Tecmer,~P.; Boguslawski,~K.; Johnson,~P.~A.; Limacher,~P.~A.; Chan,~M.;
  Verstraelen,~T.; Ayers,~P.~W. Assessing the Accuracy of New Geminal
  Approaches. \emph{J. Phys. Chem.} \textbf{2014}, \emph{A118},
  9058--9068\relax
\mciteBstWouldAddEndPuncttrue
\mciteSetBstMidEndSepPunct{\mcitedefaultmidpunct}
{\mcitedefaultendpunct}{\mcitedefaultseppunct}\relax
\EndOfBibitem
\bibitem[Boguslawski and Ayers(2015)Boguslawski, and Ayers]{boguslawski:2015}
Boguslawski,~K.; Ayers,~P.~W. Linearized Coupled Cluster Correction on the
  Antisymmetric Product of 1-Reference Orbital Geminals. \emph{J. Chem. Theory
  Comput.} \textbf{2015}, \emph{11}, 5252--5261\relax
\mciteBstWouldAddEndPuncttrue
\mciteSetBstMidEndSepPunct{\mcitedefaultmidpunct}
{\mcitedefaultendpunct}{\mcitedefaultseppunct}\relax
\EndOfBibitem
\bibitem[Boguslawski \latin{et~al.}(2016)Boguslawski, Tecmer, and
  Legeza]{boguslawski:2016a}
Boguslawski,~K.; Tecmer,~P.; Legeza,~O. Analysis of Two-Orbital Correlations in
  Wave Functions Restricted to Electron-Pair States. \emph{Phys. Rev. B}
  \textbf{2016}, \emph{94}, 155126\relax
\mciteBstWouldAddEndPuncttrue
\mciteSetBstMidEndSepPunct{\mcitedefaultmidpunct}
{\mcitedefaultendpunct}{\mcitedefaultseppunct}\relax
\EndOfBibitem
\bibitem[Boguslawski(2016)]{boguslawski:2016b}
Boguslawski,~K. Targeting Excited States in All-Trans Polyenes with
  Electron-Pair States. \emph{J. Chem. Phys.} \textbf{2016}, \emph{145},
  234105\relax
\mciteBstWouldAddEndPuncttrue
\mciteSetBstMidEndSepPunct{\mcitedefaultmidpunct}
{\mcitedefaultendpunct}{\mcitedefaultseppunct}\relax
\EndOfBibitem
\bibitem[Boguslawski and Tecmer(2017)Boguslawski, and Tecmer]{boguslawski:2017}
Boguslawski,~K.; Tecmer,~P. Benchmark of Dynamic Electron Correlation Models
  for Seniority-Zero Wave Functions and their Application to Thermochemistry.
  \emph{J. Chem. Theory Comput.} \textbf{2017}, \emph{13}, 5966--5983\relax
\mciteBstWouldAddEndPuncttrue
\mciteSetBstMidEndSepPunct{\mcitedefaultmidpunct}
{\mcitedefaultendpunct}{\mcitedefaultseppunct}\relax
\EndOfBibitem
\bibitem[Boguslawski(2019)]{boguslawski:2019}
Boguslawski,~K. Targeting Doubly Excited States with Equation of Motion Coupled
  Cluster Theory Restricted to Double Excitations. \emph{J. Chem. Theory
  Comput.} \textbf{2019}, \emph{15}, 18--24\relax
\mciteBstWouldAddEndPuncttrue
\mciteSetBstMidEndSepPunct{\mcitedefaultmidpunct}
{\mcitedefaultendpunct}{\mcitedefaultseppunct}\relax
\EndOfBibitem
\bibitem[Nowak \latin{et~al.}(2019)Nowak, Tecmer, and Boguslawski]{nowak:2019}
Nowak,~A.; Tecmer,~P.; Boguslawski,~K. Assessing the Accuracy of Simplified
  Coupled Cluster Methods for Electronic Excited State in f0 Actinide
  Compounds. \emph{Phys. Chem. Chem. Phys.} \textbf{2019}, \emph{21},
  19039--19053\relax
\mciteBstWouldAddEndPuncttrue
\mciteSetBstMidEndSepPunct{\mcitedefaultmidpunct}
{\mcitedefaultendpunct}{\mcitedefaultseppunct}\relax
\EndOfBibitem
\bibitem[Boguslawski(2021)]{boguslawski:2021}
Boguslawski,~K. Open-Shell Extensions to Closed-Shell pCCD. \emph{Chem.
  Commun.} \textbf{2021}, \emph{57}, 12277--12280\relax
\mciteBstWouldAddEndPuncttrue
\mciteSetBstMidEndSepPunct{\mcitedefaultmidpunct}
{\mcitedefaultendpunct}{\mcitedefaultseppunct}\relax
\EndOfBibitem
\bibitem[Nowak \latin{et~al.}(2021)Nowak, Legeza, and Boguslawski]{nowak:2021}
Nowak,~A.; Legeza,~O.; Boguslawski,~K. Orbital Entanglement and Correlation
  from pCCD-Tailored Coupled Cluster Wave Functions. \emph{J. Chem. Phys.}
  \textbf{2021}, \emph{154}, 084111\relax
\mciteBstWouldAddEndPuncttrue
\mciteSetBstMidEndSepPunct{\mcitedefaultmidpunct}
{\mcitedefaultendpunct}{\mcitedefaultseppunct}\relax
\EndOfBibitem
\bibitem[Marie \latin{et~al.}(2021)Marie, Kossoski, and Loos]{marie:2021}
Marie,~A.; Kossoski,~F.; Loos,~P.-F. Variational Coupled Cluster for Ground and
  Excited States. \emph{J. Chem. Phys.} \textbf{2021}, \emph{155}, 104105\relax
\mciteBstWouldAddEndPuncttrue
\mciteSetBstMidEndSepPunct{\mcitedefaultmidpunct}
{\mcitedefaultendpunct}{\mcitedefaultseppunct}\relax
\EndOfBibitem
\bibitem[Kossoski \latin{et~al.}(2021)Kossoski, Marie, Scemama, Caffarel, and
  Loos]{kossoski:2021}
Kossoski,~F.; Marie,~A.; Scemama,~A.; Caffarel,~M.; Loos,~P.-F. Excited States
  from State-Specific Orbital-Optimized Pair Coupled Cluster. \emph{J. Chem.
  Theory Comput.} \textbf{2021}, \emph{17}, 4756--4768\relax
\mciteBstWouldAddEndPuncttrue
\mciteSetBstMidEndSepPunct{\mcitedefaultmidpunct}
{\mcitedefaultendpunct}{\mcitedefaultseppunct}\relax
\EndOfBibitem
\bibitem[Bardeen \latin{et~al.}(1957)Bardeen, Cooper, and
  Schrieffer]{bardeen:1957a}
Bardeen,~J.; Cooper,~L.~N.; Schrieffer,~J.~R. Microscopic Theory of
  Superconductivity. \emph{Phys. Rev.} \textbf{1957}, \emph{106},
  162--164\relax
\mciteBstWouldAddEndPuncttrue
\mciteSetBstMidEndSepPunct{\mcitedefaultmidpunct}
{\mcitedefaultendpunct}{\mcitedefaultseppunct}\relax
\EndOfBibitem
\bibitem[Bardeen \latin{et~al.}(1957)Bardeen, Cooper, and
  Schrieffer]{bardeen:1957b}
Bardeen,~J.; Cooper,~L.~N.; Schrieffer,~J.~R. Theory of Superconductivity.
  \emph{Phys. Rev.} \textbf{1957}, \emph{108}, 1175--1204\relax
\mciteBstWouldAddEndPuncttrue
\mciteSetBstMidEndSepPunct{\mcitedefaultmidpunct}
{\mcitedefaultendpunct}{\mcitedefaultseppunct}\relax
\EndOfBibitem
\bibitem[Schrieffer(1964)]{schrieffer_book}
Schrieffer,~J.~R. \emph{Theory of Superconductivity}; CRC Press: Boca Raton,
  1964\relax
\mciteBstWouldAddEndPuncttrue
\mciteSetBstMidEndSepPunct{\mcitedefaultmidpunct}
{\mcitedefaultendpunct}{\mcitedefaultseppunct}\relax
\EndOfBibitem
\bibitem[Richardson(1963)]{richardson:1963}
Richardson,~R.~W. A Restricted Class of Exact Eigenstates of the Pairing-Force
  Hamiltonian. \emph{Phys. Lett.} \textbf{1963}, \emph{3}, 277--279\relax
\mciteBstWouldAddEndPuncttrue
\mciteSetBstMidEndSepPunct{\mcitedefaultmidpunct}
{\mcitedefaultendpunct}{\mcitedefaultseppunct}\relax
\EndOfBibitem
\bibitem[Richardson and Sherman(1964)Richardson, and Sherman]{richardson:1964}
Richardson,~R.~W.; Sherman,~N. Exact Eigenstates of the Pairing-Force
  Hamiltonian. \emph{Nucl. Phys.} \textbf{1964}, \emph{52}, 221--238\relax
\mciteBstWouldAddEndPuncttrue
\mciteSetBstMidEndSepPunct{\mcitedefaultmidpunct}
{\mcitedefaultendpunct}{\mcitedefaultseppunct}\relax
\EndOfBibitem
\bibitem[Richardson(1965)]{richardson:1965}
Richardson,~R.~W. Exact Eigenstates of the Pairing-Force Hamiltonian. II.
  \emph{J. Math. Phys.} \textbf{1965}, \emph{6}, 1034--1051\relax
\mciteBstWouldAddEndPuncttrue
\mciteSetBstMidEndSepPunct{\mcitedefaultmidpunct}
{\mcitedefaultendpunct}{\mcitedefaultseppunct}\relax
\EndOfBibitem
\bibitem[Gaudin(1976)]{gaudin:1976}
Gaudin,~M. Diagonalization of a Class of Spin Hamiltonians. \emph{J. Phys. II}
  \textbf{1976}, \emph{37}, 1087--1098\relax
\mciteBstWouldAddEndPuncttrue
\mciteSetBstMidEndSepPunct{\mcitedefaultmidpunct}
{\mcitedefaultendpunct}{\mcitedefaultseppunct}\relax
\EndOfBibitem
\bibitem[Johnson \latin{et~al.}(2020)Johnson, Fecteau, Berthiaume, Cloutier,
  Carrier, Gratton, Bultinck, De~Baerdemacker, Van~Neck, Limacher, and
  Ayers]{johnson:2020}
Johnson,~P.~A.; Fecteau,~C.-E.; Berthiaume,~F.; Cloutier,~S.; Carrier,~L.;
  Gratton,~M.; Bultinck,~P.; De~Baerdemacker,~S.; Van~Neck,~D.; Limacher,~P.
  \latin{et~al.}  Richardson-Gaudin Mean-Field for Strong Correlation in
  Quantum Chemistry. \emph{J. Chem. Phys.} \textbf{2020}, \emph{153},
  104110\relax
\mciteBstWouldAddEndPuncttrue
\mciteSetBstMidEndSepPunct{\mcitedefaultmidpunct}
{\mcitedefaultendpunct}{\mcitedefaultseppunct}\relax
\EndOfBibitem
\bibitem[Fecteau \latin{et~al.}(2022)Fecteau, Cloutier, Moisset, Boulay,
  Bultinck, Faribault, and Johnson]{fecteau:2022}
Fecteau,~C.-E.; Cloutier,~S.; Moisset,~J.-D.; Boulay,~J.; Bultinck,~P.;
  Faribault,~A.; Johnson,~P.~A. Near-Exact Treatment of Seniority-Zero Ground
  and Excited States with a Richardson-Gaudin Mean-Field. \emph{J. Chem. Phys.}
  \textbf{2022}, \emph{156}, 194103\relax
\mciteBstWouldAddEndPuncttrue
\mciteSetBstMidEndSepPunct{\mcitedefaultmidpunct}
{\mcitedefaultendpunct}{\mcitedefaultseppunct}\relax
\EndOfBibitem
\bibitem[Johnson and DePrince~III(2023)Johnson, and DePrince~III]{johnson:2023}
Johnson,~P.~A.; DePrince~III,~A.~E. Single reference treatment of strongly
  correlated H$_4$ and H$_10$ isomers with Richardson-Gaudin states. \emph{J.
  Chem. Theory Comput.} \textbf{2023}, \emph{19}, 8129--8146\relax
\mciteBstWouldAddEndPuncttrue
\mciteSetBstMidEndSepPunct{\mcitedefaultmidpunct}
{\mcitedefaultendpunct}{\mcitedefaultseppunct}\relax
\EndOfBibitem
\bibitem[Gorohovsky and Bettelheim(2011)Gorohovsky, and
  Bettelheim]{gorohovsky:2011}
Gorohovsky,~G.; Bettelheim,~E. Exact Expectation Values Within Richardson's
  Approach for the Pairing Hamiltonian in a Macroscopic System. \emph{Phys.
  Rev. B} \textbf{2011}, \emph{84}, 224503\relax
\mciteBstWouldAddEndPuncttrue
\mciteSetBstMidEndSepPunct{\mcitedefaultmidpunct}
{\mcitedefaultendpunct}{\mcitedefaultseppunct}\relax
\EndOfBibitem
\bibitem[Fecteau \latin{et~al.}(2020)Fecteau, Fortin, Cloutier, and
  Johnson]{fecteau:2020}
Fecteau,~C.-E.; Fortin,~H.; Cloutier,~S.; Johnson,~P.~A. Reduced Density
  Matrices of Richardson-Gaudin States in the Gaudin Algebra Basis. \emph{J.
  Chem. Phys.} \textbf{2020}, \emph{153}, 164117\relax
\mciteBstWouldAddEndPuncttrue
\mciteSetBstMidEndSepPunct{\mcitedefaultmidpunct}
{\mcitedefaultendpunct}{\mcitedefaultseppunct}\relax
\EndOfBibitem
\bibitem[Faribault \latin{et~al.}(2022)Faribault, Dimo, Moisset, and
  Johnson]{faribault:2022}
Faribault,~A.; Dimo,~C.; Moisset,~J.-D.; Johnson,~P.~A. Reduced Density
  Matrices/Static Correlation Functions of Richardson-Gaudin States Without
  Rapidities. \emph{J. Chem. Phys.} \textbf{2022}, \emph{157}, 214104\relax
\mciteBstWouldAddEndPuncttrue
\mciteSetBstMidEndSepPunct{\mcitedefaultmidpunct}
{\mcitedefaultendpunct}{\mcitedefaultseppunct}\relax
\EndOfBibitem
\bibitem[Johnson \latin{et~al.}(2021)Johnson, Fortin, Cloutier, and
  Fecteau]{johnson:2021}
Johnson,~P.~A.; Fortin,~H.; Cloutier,~S.; Fecteau,~C.-E. Transition Density
  Matrices of Richardson-Gaudin States. \emph{J. Chem. Phys.} \textbf{2021},
  \emph{154}, 124125\relax
\mciteBstWouldAddEndPuncttrue
\mciteSetBstMidEndSepPunct{\mcitedefaultmidpunct}
{\mcitedefaultendpunct}{\mcitedefaultseppunct}\relax
\EndOfBibitem
\bibitem[Chen and Scuseria(2023)Chen, and Scuseria]{chen:2023}
Chen,~G.~P.; Scuseria,~G.~E. Robust formulation of Wick's theorem for computing
  matrix elements between Hartree-Fock-Bogoliubov wavefunctions. \emph{J. Chem.
  Phys.} \textbf{2023}, \emph{158}, 231102\relax
\mciteBstWouldAddEndPuncttrue
\mciteSetBstMidEndSepPunct{\mcitedefaultmidpunct}
{\mcitedefaultendpunct}{\mcitedefaultseppunct}\relax
\EndOfBibitem
\bibitem[Johnson(2023)]{johnson:2023b}
Johnson,~P.~A. Richardson-Gaudin states. 2023; arXiv:2312.08804\relax
\mciteBstWouldAddEndPuncttrue
\mciteSetBstMidEndSepPunct{\mcitedefaultmidpunct}
{\mcitedefaultendpunct}{\mcitedefaultseppunct}\relax
\EndOfBibitem
\bibitem[Rombouts \latin{et~al.}(2004)Rombouts, Van~Neck, and
  Dukelsky]{rombouts:2004}
Rombouts,~S.; Van~Neck,~D.; Dukelsky,~J. Solving the Richardson Equations for
  Fermions. \emph{Phys. Rev. C} \textbf{2004}, \emph{69}, 061303(R)\relax
\mciteBstWouldAddEndPuncttrue
\mciteSetBstMidEndSepPunct{\mcitedefaultmidpunct}
{\mcitedefaultendpunct}{\mcitedefaultseppunct}\relax
\EndOfBibitem
\bibitem[Guan \latin{et~al.}(2012)Guan, Launey, Xie, Bao, Pan, and
  Draayer]{guan:2012}
Guan,~X.; Launey,~K.~D.; Xie,~M.; Bao,~L.; Pan,~F.; Draayer,~J.~P.
  Heine-Stieltjes Correspondence and the Polynomial Approach to the Standard
  Pairing Problem. \emph{Phys. Rev. C} \textbf{2012}, \emph{86}, 024313\relax
\mciteBstWouldAddEndPuncttrue
\mciteSetBstMidEndSepPunct{\mcitedefaultmidpunct}
{\mcitedefaultendpunct}{\mcitedefaultseppunct}\relax
\EndOfBibitem
\bibitem[Pogosov(2012)]{pogosov:2012}
Pogosov,~W.~V. `Probabilistic' Approach to Richardson Equations. \emph{J. Phys.
  Condens. Matter} \textbf{2012}, \emph{24}, 075701\relax
\mciteBstWouldAddEndPuncttrue
\mciteSetBstMidEndSepPunct{\mcitedefaultmidpunct}
{\mcitedefaultendpunct}{\mcitedefaultseppunct}\relax
\EndOfBibitem
\bibitem[De~Baerdemacker(2012)]{debaerdemacker:2012}
De~Baerdemacker,~S. Richardson-Gaudin Integrability in the Contraction Limit of
  the Quasispin. \emph{Phys. Rev. C} \textbf{2012}, \emph{86}, 044332\relax
\mciteBstWouldAddEndPuncttrue
\mciteSetBstMidEndSepPunct{\mcitedefaultmidpunct}
{\mcitedefaultendpunct}{\mcitedefaultseppunct}\relax
\EndOfBibitem
\bibitem[Faribault \latin{et~al.}(2011)Faribault, El~Araby, Str\"ater, and
  Gritsev]{faribault:2011}
Faribault,~A.; El~Araby,~O.; Str\"ater,~C.; Gritsev,~V. Gaudin Models Solver
  Based on the Correspondence Between Bethe Ansatz and Ordinary Differential
  Equations. \emph{Phys. Rev. B} \textbf{2011}, \emph{83}, 235124\relax
\mciteBstWouldAddEndPuncttrue
\mciteSetBstMidEndSepPunct{\mcitedefaultmidpunct}
{\mcitedefaultendpunct}{\mcitedefaultseppunct}\relax
\EndOfBibitem
\bibitem[El~Araby \latin{et~al.}(2012)El~Araby, Gritsev, and
  Faribault]{elaraby:2012}
El~Araby,~O.; Gritsev,~V.; Faribault,~A. \emph{Phys. Rev. B} \textbf{2012},
  \emph{85}, 115130\relax
\mciteBstWouldAddEndPuncttrue
\mciteSetBstMidEndSepPunct{\mcitedefaultmidpunct}
{\mcitedefaultendpunct}{\mcitedefaultseppunct}\relax
\EndOfBibitem
\bibitem[Yuzbashyan \latin{et~al.}(2003)Yuzbashyan, Baytin, and
  Altshuler]{yuz:2003}
Yuzbashyan,~E.~A.; Baytin,~A.~A.; Altshuler,~B.~L. Strong-coupling expansion
  for the pairing Hamiltonian for small superconducting metallic grains.
  \emph{Phys. Rev. B} \textbf{2003}, \emph{68}, 214509\relax
\mciteBstWouldAddEndPuncttrue
\mciteSetBstMidEndSepPunct{\mcitedefaultmidpunct}
{\mcitedefaultendpunct}{\mcitedefaultseppunct}\relax
\EndOfBibitem
\bibitem[Yuzbashyan \latin{et~al.}(2005)Yuzbashyan, Baytin, and
  Altshuler]{yuz:2005}
Yuzbashyan,~E.~A.; Baytin,~A.~A.; Altshuler,~B.~L. Finite-size corrections for
  the pairing Hamiltonian. \emph{Phys. Rev. B} \textbf{2005}, \emph{71},
  094505\relax
\mciteBstWouldAddEndPuncttrue
\mciteSetBstMidEndSepPunct{\mcitedefaultmidpunct}
{\mcitedefaultendpunct}{\mcitedefaultseppunct}\relax
\EndOfBibitem
\bibitem[Yuzbashyan \latin{et~al.}(2002)Yuzbashyan, Altshuler, and
  Shastry]{yuz:2002}
Yuzbashyan,~E.~A.; Altshuler,~B.~L.; Shastry,~B.~S. The origin of degeneracies
  and crossings in the 1d Hubbard model. \emph{J. Phys. A Math. Gen.}
  \textbf{2002}, \emph{35}, 7525\relax
\mciteBstWouldAddEndPuncttrue
\mciteSetBstMidEndSepPunct{\mcitedefaultmidpunct}
{\mcitedefaultendpunct}{\mcitedefaultseppunct}\relax
\EndOfBibitem
\bibitem[Sklyanin(1989)]{sklyanin:1989}
Sklyanin,~E.~K. Separation of Variables in the Gaudin Model. \emph{J. Sov.
  Math.} \textbf{1989}, \emph{47}, 2473--2488\relax
\mciteBstWouldAddEndPuncttrue
\mciteSetBstMidEndSepPunct{\mcitedefaultmidpunct}
{\mcitedefaultendpunct}{\mcitedefaultseppunct}\relax
\EndOfBibitem
\bibitem[Cambiaggio \latin{et~al.}(1997)Cambiaggio, Rivas, and
  Saraceno]{cambiaggio:1997}
Cambiaggio,~M.~C.; Rivas,~A. M.~F.; Saraceno,~M. Integrability of the Pairing
  Hamiltonian. \emph{Nucl. Phys. A} \textbf{1997}, \emph{624}, 157--167\relax
\mciteBstWouldAddEndPuncttrue
\mciteSetBstMidEndSepPunct{\mcitedefaultmidpunct}
{\mcitedefaultendpunct}{\mcitedefaultseppunct}\relax
\EndOfBibitem
\bibitem[Faribault \latin{et~al.}(2008)Faribault, Calabrese, and
  Caux]{faribault:2008}
Faribault,~A.; Calabrese,~P.; Caux,~J.-S. Exact Mesoscopic Correlation
  Functions of the Richardson Pairing Model. \emph{Phys. Rev. B} \textbf{2008},
  \emph{77}, 064503\relax
\mciteBstWouldAddEndPuncttrue
\mciteSetBstMidEndSepPunct{\mcitedefaultmidpunct}
{\mcitedefaultendpunct}{\mcitedefaultseppunct}\relax
\EndOfBibitem
\bibitem[Faribault \latin{et~al.}(2010)Faribault, Calabrese, and
  Caux]{faribault:2010}
Faribault,~A.; Calabrese,~P.; Caux,~J.-S. Dynamical Correlation Functions of
  the Mesoscopic Pairing Model. \emph{Phys. Rev. B} \textbf{2010}, \emph{81},
  174507\relax
\mciteBstWouldAddEndPuncttrue
\mciteSetBstMidEndSepPunct{\mcitedefaultmidpunct}
{\mcitedefaultendpunct}{\mcitedefaultseppunct}\relax
\EndOfBibitem
\bibitem[Vein and Dale(1999)Vein, and Dale]{vein_book}
Vein,~R.; Dale,~P. \emph{Determinants and Their Applications in Mathematical
  Physics}; Springer-Verlag: New York, 1999\relax
\mciteBstWouldAddEndPuncttrue
\mciteSetBstMidEndSepPunct{\mcitedefaultmidpunct}
{\mcitedefaultendpunct}{\mcitedefaultseppunct}\relax
\EndOfBibitem
\bibitem[Claeys \latin{et~al.}(2017)Claeys, Van~Neck, and
  De~Baerdemacker]{claeys:2017b}
Claeys,~P.~W.; Van~Neck,~D.; De~Baerdemacker,~S. Inner Products in Integrable
  Richardson-Gaudin Models. \emph{SciPost Phys.} \textbf{2017}, \emph{3},
  028\relax
\mciteBstWouldAddEndPuncttrue
\mciteSetBstMidEndSepPunct{\mcitedefaultmidpunct}
{\mcitedefaultendpunct}{\mcitedefaultseppunct}\relax
\EndOfBibitem
\bibitem[Goddard~III(1967)]{goddard:1967}
Goddard~III,~W.~A. Improved Quantum Theory of Many-Electron Systems. II. The
  Basic Method. \emph{Phys. Rev.} \textbf{1967}, \emph{157}, 81\relax
\mciteBstWouldAddEndPuncttrue
\mciteSetBstMidEndSepPunct{\mcitedefaultmidpunct}
{\mcitedefaultendpunct}{\mcitedefaultseppunct}\relax
\EndOfBibitem
\bibitem[Hay \latin{et~al.}(1972)Hay, Hunt, and Goddard~III]{hay:1972}
Hay,~P.~J.; Hunt,~W.~J.; Goddard~III,~W.~A. Generalized Valence Bond
  Wavefunctions for the Low Lying Excited States of Methylene. \emph{Chem.
  Phys. Lett.} \textbf{1972}, \emph{13}, 30--35\relax
\mciteBstWouldAddEndPuncttrue
\mciteSetBstMidEndSepPunct{\mcitedefaultmidpunct}
{\mcitedefaultendpunct}{\mcitedefaultseppunct}\relax
\EndOfBibitem
\bibitem[Hunt \latin{et~al.}(1972)Hunt, Hay, and Goddard~III]{hunt:1972}
Hunt,~W.~J.; Hay,~P.~J.; Goddard~III,~W.~A. Self-Consistent Procedures for
  Generalized Valence Bond Wavefunctions. Applications H$_3$, BH, H$_2$O,
  C$_2$H$_6$, and O$_2$. \emph{J. Chem. Phys.} \textbf{1972}, \emph{57},
  738--748\relax
\mciteBstWouldAddEndPuncttrue
\mciteSetBstMidEndSepPunct{\mcitedefaultmidpunct}
{\mcitedefaultendpunct}{\mcitedefaultseppunct}\relax
\EndOfBibitem
\bibitem[Goddard~III \latin{et~al.}(1973)Goddard~III, Dunning, Hunt, and
  Hay]{goddard:1973}
Goddard~III,~W.~A.; Dunning,~T.~H.; Hunt,~W.~J.; Hay,~P.~J. Generalized Valence
  Bond Description of Bonding in Low-Lying States of Molecules. \emph{Acc.
  Chem. Res.} \textbf{1973}, \emph{6}, 368--376\relax
\mciteBstWouldAddEndPuncttrue
\mciteSetBstMidEndSepPunct{\mcitedefaultmidpunct}
{\mcitedefaultendpunct}{\mcitedefaultseppunct}\relax
\EndOfBibitem
\bibitem[Goddard~III and Harding(1978)Goddard~III, and Harding]{goddard:1978}
Goddard~III,~W.~A.; Harding,~L.~B. The Description of Chemical Bonding from Ab
  Initio Calculations. \emph{Annu. Rev. Phys. Chem.} \textbf{1978}, \emph{29},
  363--396\relax
\mciteBstWouldAddEndPuncttrue
\mciteSetBstMidEndSepPunct{\mcitedefaultmidpunct}
{\mcitedefaultendpunct}{\mcitedefaultseppunct}\relax
\EndOfBibitem
\bibitem[Beran \latin{et~al.}(2005)Beran, Austin, Sodt, and
  Head-Gordon]{beran:2005}
Beran,~G. J.~O.; Austin,~B.; Sodt,~A.; Head-Gordon,~M. Unrestricted perfect
  pairing: The simplest wave-function-based model chemistry beyond mean field.
  \emph{J. Phys. Chem. A} \textbf{2005}, \emph{109}, 9183--9192\relax
\mciteBstWouldAddEndPuncttrue
\mciteSetBstMidEndSepPunct{\mcitedefaultmidpunct}
{\mcitedefaultendpunct}{\mcitedefaultseppunct}\relax
\EndOfBibitem
\bibitem[Small and Head-Gordon(2009)Small, and Head-Gordon]{small:2009}
Small,~D.~W.; Head-Gordon,~M. Tractable spin-pure methods for bond breaking:
  Local many-electron spin-vector sets and an approximate valence bond model.
  \emph{J. Chem. Phys.} \textbf{2009}, \emph{130}, 084103\relax
\mciteBstWouldAddEndPuncttrue
\mciteSetBstMidEndSepPunct{\mcitedefaultmidpunct}
{\mcitedefaultendpunct}{\mcitedefaultseppunct}\relax
\EndOfBibitem
\bibitem[Bytautas \latin{et~al.}(2011)Bytautas, Henderson, Jimenez-Hoyos,
  Ellis, and Scuseria]{bytautas:2011}
Bytautas,~L.; Henderson,~T.~M.; Jimenez-Hoyos,~C.~A.; Ellis,~J.~K.;
  Scuseria,~G.~E. Seniority and Orbital Symmetry as Tools for Establishing a
  Full Configuration Interaction Hierarchy. \emph{J. Chem. Phys.}
  \textbf{2011}, \emph{135}, 044119\relax
\mciteBstWouldAddEndPuncttrue
\mciteSetBstMidEndSepPunct{\mcitedefaultmidpunct}
{\mcitedefaultendpunct}{\mcitedefaultseppunct}\relax
\EndOfBibitem
\bibitem[Carrier \latin{et~al.}(2020)Carrier, Fecteau, and
  Johnson]{carrier:2020}
Carrier,~L.; Fecteau,~C.-E.; Johnson,~P.~A. Bethe Ansatz of Electrons as a
  Mean-Field Wavefunction for Chemical Systems. \emph{Int. J. Quantum Chem.}
  \textbf{2020}, \emph{120}, e26255\relax
\mciteBstWouldAddEndPuncttrue
\mciteSetBstMidEndSepPunct{\mcitedefaultmidpunct}
{\mcitedefaultendpunct}{\mcitedefaultseppunct}\relax
\EndOfBibitem
\bibitem[Moisset \latin{et~al.}(2022)Moisset, Carrier, and
  Johnson]{moisset:2022b}
Moisset,~J.-D.; Carrier,~L.; Johnson,~P.~A. Perturbative Corrections for
  Hartree-Fock-Like Algebraic Bethe Ansatz Analogue. \emph{J. Math. Chem.}
  \textbf{2022}, \emph{60}, 1707--1724\relax
\mciteBstWouldAddEndPuncttrue
\mciteSetBstMidEndSepPunct{\mcitedefaultmidpunct}
{\mcitedefaultendpunct}{\mcitedefaultseppunct}\relax
\EndOfBibitem
\bibitem[Epstein(1926)]{epstein:1926}
Epstein,~P.~S. The Stark Effect from the Point of View of Schroedinger's
  Quantum Theory. \emph{Phys. Rev.} \textbf{1926}, \emph{28}, 695\relax
\mciteBstWouldAddEndPuncttrue
\mciteSetBstMidEndSepPunct{\mcitedefaultmidpunct}
{\mcitedefaultendpunct}{\mcitedefaultseppunct}\relax
\EndOfBibitem
\bibitem[Nesbet(1955)]{nesbet:1955}
Nesbet,~R.~K. Configuration interaction in orbital theories. \emph{Proc. R.
  Soc. Lond. Ser. A} \textbf{1955}, \emph{230}, 312--321\relax
\mciteBstWouldAddEndPuncttrue
\mciteSetBstMidEndSepPunct{\mcitedefaultmidpunct}
{\mcitedefaultendpunct}{\mcitedefaultseppunct}\relax
\EndOfBibitem
\bibitem[Stair and Evangelista(2020)Stair, and Evangelista]{stair:2020}
Stair,~N.~H.; Evangelista,~F.~A. Exploring Hilbert Space on a Budget: Novel
  Benchmark Set and Performance Metric for Testing Electronic Structure Methods
  in the Regime of Strong Correlation. \emph{J. Chem. Phys.} \textbf{2020},
  \emph{153}, 104108\relax
\mciteBstWouldAddEndPuncttrue
\mciteSetBstMidEndSepPunct{\mcitedefaultmidpunct}
{\mcitedefaultendpunct}{\mcitedefaultseppunct}\relax
\EndOfBibitem
\bibitem[Piris(2017)]{piris:2017}
Piris,~M. Global Method for Electron Correlation. \emph{Phys. Rev. Lett.}
  \textbf{2017}, \emph{119}, 063002\relax
\mciteBstWouldAddEndPuncttrue
\mciteSetBstMidEndSepPunct{\mcitedefaultmidpunct}
{\mcitedefaultendpunct}{\mcitedefaultseppunct}\relax
\EndOfBibitem
\bibitem[Mitxelena and Piris(2020)Mitxelena, and Piris]{mitxelena:2020a}
Mitxelena,~I.; Piris,~M. An Efficient Method for Strongly Correlated Electrons
  in One Dimension. \emph{J. Phys. Condens. Matter} \textbf{2020}, \emph{32},
  17LT01\relax
\mciteBstWouldAddEndPuncttrue
\mciteSetBstMidEndSepPunct{\mcitedefaultmidpunct}
{\mcitedefaultendpunct}{\mcitedefaultseppunct}\relax
\EndOfBibitem
\bibitem[Mitxelena and Piris(2020)Mitxelena, and Piris]{mitxelena:2020b}
Mitxelena,~I.; Piris,~M. An Efficient Method for Strongly Correlated Electrons
  in Two-Dimensions. \emph{J. Chem. Phys.} \textbf{2020}, \emph{152},
  064108\relax
\mciteBstWouldAddEndPuncttrue
\mciteSetBstMidEndSepPunct{\mcitedefaultmidpunct}
{\mcitedefaultendpunct}{\mcitedefaultseppunct}\relax
\EndOfBibitem
\bibitem[Mitxelena and Piris(2022)Mitxelena, and Piris]{mitxelena:2022}
Mitxelena,~I.; Piris,~M. Benchmarking GNOF Against FCI in Challenging Systems
  in One, Two, and Three Dimensions. \emph{J. Chem. Phys.} \textbf{2022},
  \emph{156}, 214102\relax
\mciteBstWouldAddEndPuncttrue
\mciteSetBstMidEndSepPunct{\mcitedefaultmidpunct}
{\mcitedefaultendpunct}{\mcitedefaultseppunct}\relax
\EndOfBibitem
\bibitem[Hurley \latin{et~al.}(1953)Hurley, Lennard-Jones, and
  Pople]{hurley:1953}
Hurley,~A.~C.; Lennard-Jones,~J.~E.; Pople,~J.~A. The Molecular Orbital Theory
  of Chemical Valency XVI. A Theory of Paired-Electrons in Polyatomic
  Molecules. \emph{Proc. R. Soc.} \textbf{1953}, \emph{A220}, 446--455\relax
\mciteBstWouldAddEndPuncttrue
\mciteSetBstMidEndSepPunct{\mcitedefaultmidpunct}
{\mcitedefaultendpunct}{\mcitedefaultseppunct}\relax
\EndOfBibitem
\bibitem[Kutzelnigg(1964)]{kutzelnigg:1964}
Kutzelnigg,~W. Direct Determination of Natural Orbitals and Natural Expansion
  Coefficients of Many-Electron Wavefunctions. I. Natural Orbitals in the
  Geminal Product Approximation. \emph{J. Chem. Phys.} \textbf{1964},
  \emph{40}, 3640--3647\relax
\mciteBstWouldAddEndPuncttrue
\mciteSetBstMidEndSepPunct{\mcitedefaultmidpunct}
{\mcitedefaultendpunct}{\mcitedefaultseppunct}\relax
\EndOfBibitem
\bibitem[Nicely and Harrison(1971)Nicely, and Harrison]{nicely:1971}
Nicely,~V.~A.; Harrison,~J.~F. Geminal Product Wavefunctions: A General
  Formalism. \emph{J. Chem. Phys.} \textbf{1971}, \emph{54}, 4363--4368\relax
\mciteBstWouldAddEndPuncttrue
\mciteSetBstMidEndSepPunct{\mcitedefaultmidpunct}
{\mcitedefaultendpunct}{\mcitedefaultseppunct}\relax
\EndOfBibitem
\bibitem[Coleman(1965)]{coleman:1965}
Coleman,~A.~J. Structure of Fermion Density Matrices. II. Antisymmetrized
  Geminal Powers. \emph{J. Math. Phys.} \textbf{1965}, \emph{6},
  1425--1431\relax
\mciteBstWouldAddEndPuncttrue
\mciteSetBstMidEndSepPunct{\mcitedefaultmidpunct}
{\mcitedefaultendpunct}{\mcitedefaultseppunct}\relax
\EndOfBibitem
\bibitem[Coleman(1989)]{coleman:1989}
Coleman,~A.~J. The Structure of Fermion Density Matrices. III. Long-Range
  Order. \emph{J. Low Temp. Phys.} \textbf{1989}, \emph{74}, 1--17\relax
\mciteBstWouldAddEndPuncttrue
\mciteSetBstMidEndSepPunct{\mcitedefaultmidpunct}
{\mcitedefaultendpunct}{\mcitedefaultseppunct}\relax
\EndOfBibitem
\bibitem[Coleman(1997)]{coleman:1997}
Coleman,~A.~J. The AGP Model for Fermion Systems. \emph{Int. J. Quantum Chem.}
  \textbf{1997}, \emph{63}, 23--30\relax
\mciteBstWouldAddEndPuncttrue
\mciteSetBstMidEndSepPunct{\mcitedefaultmidpunct}
{\mcitedefaultendpunct}{\mcitedefaultseppunct}\relax
\EndOfBibitem
\end{mcitethebibliography}

\end{document}